\documentclass[12pt]{article}
\usepackage{epsfig}
\usepackage{tikz}
\usetikzlibrary{shapes,arrows}
\tikzstyle{line} = [draw, -latex']
\tikzstyle{block} = [rectangle, text width=5em, text centered, rounded corners, minimum height=4em]

\def\gtap{\raisebox{-.4ex}{\rlap{$\sim$}} \raisebox{.4ex}{$>$}} 
\def\beq{\begin{equation}} 
\def\eeq{\end{equation}} 
\def\barr{\begin{array}}
\def\earr{\end{array}}
\def\dis{\displaystyle}

\newcommand\npb[3]{{Nucl.\ Phys.\ }{\bf B #1} (#2) #3}
\newcommand\zpc[3]{{Z.\ Physik }{\bf C #1} (#2) #3}

\newcommand{\hmu}{{\hat\mu}}
\newcommand{\hnu}{{\hat\nu}}
 
\newcommand{\hh}{{\hat{h}}}

\newcommand{\hk}{{\hat\kappa}}

\newcommand{\vn}{{\vec{n}}}

\newcommand{\gsim}{\lower.7ex\hbox{$\;\stackrel{\textstyle>}{\sim}\;$}}
\newcommand{\lsim}{\lower.7ex\hbox{$\;\stackrel{\textstyle<}{\sim}\;$}}
\newcommand{\del}{\partial}

\tikzstyle{block} = [rectangle, draw, fill=blue!20, 
    text width=5em, text centered, rounded corners, minimum height=4em]

\begin{document}
\thispagestyle{empty}
\begin{flushright}
\texttt{HIP-2012-01/TH}\\

\end{flushright}
\vskip 15pt

\begin{center}
{\Large {\bf Constraints on Universal Extra Dimension models with gravity mediated decays from {\em ATLAS} diphoton search}}
\renewcommand{\thefootnote}{\alph{footnote}}

\hspace*{\fill}

\hspace*{\fill}

{ \tt{
Kirtiman Ghosh\footnote{kirtiman.ghosh@helsinki.fi}, Katri Huitu\footnote{katri.huitu@helsinki.fi} 
}}\\

\small {\em Department of Physics and Helsinki Institute of Physics\\
 FIN-00014, University of Helsinki\\ Finland.}\\

\vskip 40pt

{\bf ABSTRACT}
\end{center}
\noindent We discuss the collider phenomenology of Universal Extra Dimension models with gravity mediated decays. We concentrate on diphoton associated with large missing transverse energy signature. At the collider, level-1 Kaluza-Klein (KK) particles are produced in pairs due to the conservation of KK-parity. Subsequently, KK-particles decay via cascades involving lighter KK-particles until reaching the lightest KK-particle (LKP). Finally, gravity induced decay of the LKP into photons gives rise to the diphoton signature. The search for diphoton events with large missing transverse energy was recently communicated by the {\em ATLAS} collaboration for 7 TeV center-of-mass energy and 3.1 inverse femtobarn integrated luminosity of the Large Hadron Collider. Above the Standard Model background prediction, no excess of such events was reported. We translate the absence of any excess of the diphoton events to constrain the model parameters, namely, the radius of compactification ($R$) and the fundamental Planck mass ($M_D$). 

\vskip 30pt

\section{Introduction}
The Large Hadron Collider (LHC) at CERN is aiming to reveal the mechanism for electroweak symmetry breaking (EWSB) as well as to uncover any new dynamics that may be operative at the scale of a few TeVs.
Except for the tiny mass of the neutrinos, the Standard Model (SM) remains very successful in explaining the experimental data related to the elementary particle physics. 
There is also some tension between theoretical calculations and measurements in the flavor sector, but so far the significance of those differences is not large enough to claim new physics beyond the SM. Anyway, these hints may indicate that new physics is within reach of the LHC.
The known problems of SM, like lack of dark matter and the hierarchy problem, have motivated a number of attempts to go beyond the SM. 
In this endeavour, lots of attention have been paid to the theories with one or more extra space-like dimensions.

Extra dimensional theories can be classified into several
classes \cite{add,rs,UED}. In one class of models, gravity lives in $D=(4+N)$ dimensions
and the SM particles are confined to a 3-brane (a $(3+1)$ dimensional
space) embedded in the $(4+N)$ dimensional bulk, with $N$ spatial
dimensions compactified on a volume $V_N$. For large enough size of
this extra dimensional volume $V_N$, the fundamental $D$ dimensional
Planck mass ($M_D$) can be as low as $1$ TeV, although the effective four
dimensional Planck mass ($M_{Pl}$) can be as large as $10^{19}$ GeV. The existence
of a TeV scale Planck mass automatically solves naturalness/hierarchy
problem of the SM. Models of ADD \cite{add} 
fall in this category.

On the other hand, there are a class of models, known as 
  Universal Extra Dimension (UED) models \cite{UED,KKCD,2ued}, which have flat metric but small compactification radius of ${\cal O}({\rm TeV^{-1}})$. Moreover, in the UED models, all the SM fields can propagate in
  the the extra 
  dimension(s) or bulk. As a result of compactification, every field decomposes into an infinite tower of Kaluza-Klein (KK) modes, characterized by an integer $n$, known as the KK-number. The zero modes ($n=0$ states) are identified as the corresponding SM states. One should
note that UED models do not address the gauge hierarchy
  problem as elegantly as ADD. However, there are
  several other motivations for UED models. As for example, apart from the rich collider phenomenology \cite{collider}, UED models in
general offer possible unification of the gauge couplings at a
relatively low scale of energy, not far beyond the reach of the LHC \cite{unificUED}.  Moreover, particle spectra of
UED models naturally contain a weakly interacting stable massive
particle, which can be a good candidate for cold dark matter (CDM)
\cite{dark_ued5,dark_ued6}. Apart from these generic advantages, a
  particular variant of the UED model where the number of extra
  dimensions is two, namely the { two Universal Extra Dimension}
  (2UED) model \cite{2ued,coll2ued} has some additional attractive features. 2UED model can
  naturally 
  explain the long life time for proton decay \cite{dobrescu} and more
  interestingly it predicts that the number of fermion generations
  should be an integral multiple of three \cite{dobrescu1}.
The key feature of the { UED Lagrangian} is that the momentum in the universal extra dimensions is conserved. From a 4-dimensional perspective, this implies { KK-number} conservation. However, boundary conditions break this symmetry, leaving behind only a conserved { KK-parity}, defined as $(-1)^n$, where $n$ is the { KK-number}. This discrete symmetry ensures that the lightest KK-particle (LKP) is stable\footnote{In the framework of the { minimal UED (mUED)} model, LKP is the level-1 excitation of $B_\mu$, denoted by $\gamma_1$.} and the level-1 KK-modes would be produced only in pairs.

There are interesting generalizations of the ADD scenario in which the SM particles are confined to a $(3 + m)$-brane ($3 + m + 1$ dimensional manifold) embedded in a $(4 + N)$ dimensional bulk \cite{NPB550}. Since $m$ spatial dimensions are compact, in this framework, the effective 4-dimensional theory also contains the KK-excitations of SM fields. The volume of $m$ spatial dimensions (internal to the bulk) cannot be too large due to the experimental lower bound on the KK-mode masses. This scenario is often known as { "fat brane"} scenario because the $m$ small spatial dimensions, accessible for both matter and gravity, can be viewed as the thickness of the SM 3-brane in the $(4 + N)$-dimensional bulk \cite{PLB482}. Therefore, in this scenario gravity propagates in $N$ extra dimensions with eV$^{-1}$ size.
However, matter propagation is restricted only to a small length ($\sim$ TeV$^{-1}$) which is associated with the thickness of the SM 3-brane along these extra dimensions. From a phenomenological point of view, this scenario has very interesting consequences \cite{PRD66} at the collider experiments. In this work, we have concentrated on the phenomenology of the { "fat brane"} scenario in the context of LHC with $\sqrt s=7$ TeV. 

In the framework of the { "fat brane"} scenario, the { gravity induced} interactions do not respect { KK-number {\rm or} KK-parity} conservation. The { gravity induced} interactions allow the level-1 KK-excitations of the matter fields to decay directly into their SM partners by radiating gravity excitations. Therefore, in this scenario, LKP is no more a stable particle. LKP can decay into a photon or $Z$-boson in association with a gravity excitation. This makes the collider phenomenology of this model drastically different from the phenomenology of { UED} models without { gravity induced} decays. At the LHC, { KK-parity} conservation allows the pair production of the level-1 KK-particles. The level-1 particles produced at colliders can either decay directly to their SM partners via { gravity induced} interactions or decay into lighter level-1 matter fields via { KK-number conserving} interactions. If the { gravity induced} decays dominate over the { KK-number conserving} decays, the pair production of colored\footnote{The production cross-sections of colored level-1 particles are enhanced by the strong coupling constant and color factors. Therefore, compared to the former cross-section, the pair production cross-section of color singlet level-1 particles are suppressed by a few orders of magnitude. Therefore, in this work, we have considered only the pair production of colored level-1 particles.} level-1 particles gives rise to { di-jets} in association with { large missing transverse momentum} ($p_T\!\!\!\!\!\!/~~$) signature. However, if the { KK-number conserving} decays dominate, the KK-particles decay via cascades involving lighter KK-particles until reaching the LKP at the end of the decay chain. LKP further decays into a photon or a $Z$-boson $+$ a gravity excitation. Therefore, in this case, the pair production of level-1 particles gives rise to $\gamma\gamma~(Z\gamma~{\rm or}~ZZ)+p_T\!\!\!\!\!\!/~~X$ final state where $X$ represents jets and leptons emitted in the cascade decays. 

Recently, a search for diphoton events with large $p_T\!\!\!\!\!\!/~~$ was performed by the {\em ATLAS} collaboration \cite{ATLAS}. The search was based on the data collected with the {\em ATLAS} detector in proton-proton collisions at $\sqrt s=7$ TeV and integrated luminosity 3.1 pb$^{-1}$. No excess of diphoton events was observed above the predicted SM background. Absence of any excess over the SM background, is then translated to impose bounds on the parameters of the { "fat brane"} scenario. However, the bound obtained in Ref.~\cite{ATLAS} is valid for a particular choice of the number of { "large"} extra dimension, $N$. Moreover, in the analysis of Ref.~\cite{ATLAS}, it was assumed that the LKP decays into photon $+$ gravity excitation with 100\% branching fraction. We found that this assumption cannot be justified for the present physics scenario. Because, in this scenario, LKP also decays in to $Z+$ a gravity excitation with non-negligible branching fraction. In this paper, we have reanalyzed the results obtained in Ref.~\cite{ATLAS} for different number of { "large"} extra dimensions. We have also included the effects of  non-negligible LKP $\to Z$ + graviton branching fraction. 

The rest of the paper is organized as follows. In the next section, we briefly discuss the { "fat brane"} scenario with emphasis on the { gravity matter interactions}. In section 3, we will discuss the phenomenology of this model. Finally, we summarize in section 4.

\section{The Model}

The { minimal Universal Extra Dimension (mUED)} scenario is an extension of the SM in which all particles, fermions as well as bosons, propagate into a single ${\rm TeV}^{-1}$-size extra compact dimension. The { mUED} scenario could be potentially embedded in a larger space { i.e.}, ($4+N$)-dimensional space, where only gravity propagates in the $N-1$ large extra dimensions. In this section, we will discuss one such scenario.

\subsection{{The Minimal Universal Extra Dimension} model}

In the { minimal} version of { UED (mUED)}, there is only one extra
dimension, $y$, compactified on a circle of radius $R$ with a $Z_2$ orbifolding
defined by identifying $y \to -y$. The orbifolding is crucial in generating
{chiral} zero modes for fermions. Each component of a
5-dimensional field must be either even or odd under the orbifold
projection. The $Z_2$ symmetry breaks the translational 
invariance along the $5$th dimension and generates two 
{ fixed points} at $y=0$ and $y=\pi R$. The size of 
the extra dimension is taken to be small enough 
so that one can dimensionally reduce the theory and 
construct the effective 4D Lagrangian. 
The low energy 
effective  
Lagrangian contains infinite number of Kaluza-Klein (KK) excitations 
(identified by an integer number $n$, called the KK-number)
for all the fields which are present in the 
higher dimensional Lagrangian.
 KK-mode expansions of 
different fields are presented in Appendix~\ref{KKEXP}. 

One of the interesting feature of the { mUED} model is the conservation of the KK-number.
Since all particles can propagate in the extra dimension, the momentum along the extra dimension 
is conserved and it is also quantized because of the compactification of the extra dimension $y$. 
The five dimensional momentum conservation is translated into the conservation of the KK-number 
in the four dimensions.
 However, the presence of  
two fixed points break the translational symmetry and the KK-number is not a good quantum number. 
In principle, there may exist some 
operators located at these fixed points and 
one can expect mixing among different KK-states. However, if the localized operators are 
symmetric under the exchange\footnote{This is another $Z_2$ symmetry, but 
not the $Z_2$ of $y \leftrightarrow$ $-y$.} of the fixed points, the conservation of the KK-number breaks down to the conservation 
of the KK-parity defined as $(- 1)^n$, where $n$ is the KK-number. 
The conservation of the KK-parity ensures that $n=1$ particles are always produced in pairs 
and the lightest $n=1$ particle (LKP) must be stable. It also forbids tree level mUED contribution to 
any SM process. The situation is analogous
to the $R$ parity conserving supersymmetric models \cite{susy}.

The tree-level mass of a level-$n$ KK-particle is given by $m_n^2 =
m_0^2 + {n^2}/{R^2}$, where $m_0$ is the mass associated with the
corresponding SM field. Therefore, the tree level mUED spectrum is
very nearly degenerate and, to start with, the first excitation of any
massless SM particle can be the LKP. In
practice, radiative corrections \cite{Cheng:2002iz} play an important
role in determining the actual spectrum. The correction term can be
finite (bulk correction) or it may depend on $\Lambda$, the cut-off
scale of the model (boundary correction). Bulk corrections arise due
to the winding of the internal lines in a loop around the compactified
direction~\cite{Cheng:2002iz}, and are nonzero and finite only for the
gauge boson KK-excitations. On the other hand, the boundary
corrections are not finite, but are logarithmically divergent. The bulk and
boundary corrections for level-n doublet quarks and leptons ($Q_n$ and
$L_n$), singlet quarks and leptons ($q_n$ and $e_n$) and KK-gauge
bosons ($g_n$, $W_n$, $Z_n$ and $B_n$) are presented 
in Appendix~\ref{correction}.

The KK-excitations of the neutral electroweak gauge bosons mix in a
fashion analogous to their SM counterparts and the mass eigenstates
and eigenvalues of the KK "photons" and "$Z$" bosons are obtained by
diagonalizing the corresponding mass squared matrices.  In the $(B_n,
W^3_n)$ basis, the latter reads 
\[
\left( \barr{cc} 
\dis
\frac{n^2}{R^2}+ \hat{\delta} m_{B_n}^2 + \frac{1}{4}g_1^2 v^2 
& \dis
\frac{1}{4}g_1 g_2 v^2 
\\[2ex] 
\dis 
\frac{1}{4}g_1 g_2 v^2 & \dis
\frac{n^2}{R^2}+ \hat{\delta} m_{W_n}^2 +\frac{1}{4}g_2^2 v^2 \earr
\right), 
\]
where $\hat{\delta}$ represents the total
one-loop correction, including both bulk and boundary
contributions (see Appendix~\ref{correction}). Note that, with $v$ being just the scale of EWSB, the
extent of mixing is miniscule even at $R^{-1} = 500$ GeV and is
progressively smaller for the higher KK-modes. As a consequence, unless 
$R^{-1}$ is very small, the  $Z_{1}$ and $\gamma_1$ are, 
for all practical purposes, essentially 
$W_1^3$ and $B_1$. This has profound
consequences in the decays of the KK-excitations.

These radiative corrections partially remove the degeneracy in the
spectrum \cite{Cheng:2002iz} and, over most of the parameter space,
$\gamma_1$, the first excitation of the hypercharge gauge boson ($B$),
is the LKP.  The $\gamma_1$ can
produce the right amount of relic density and turns out to be a good
dark matter candidate \cite{dark_ued5}. The mass of $\gamma_1$ is
approximately $R^{-1}$ and hence the overclosure of the universe puts
an upper bound on $R^{-1} < $ 1400 GeV. The lower limit on $R^{-1}$
comes from the low energy observables and direct search of new
particles at the Tevatron. Constraints from $g-2$ of the muon
\cite{nath}, flavour changing neutral currents \cite{chk,buras,desh},
$Z \to b\bar{b}$ \cite{santa}, the $\rho$ parameter
\cite{appel-yee}, other electroweak precision tests \cite{ewued},
etc. imply that $R^{-1}~\gtap~300$ GeV. The masses of KK-particles are
also dependent on $\Lambda$, the cut-off of UED as an effective
theory, which is essentially a free parameter. One loop corrected
$SU(3)$, $SU(2)$ and $U(1)$ gauge couplings show power law running in
the mUED model and almost meet at the scale $\Lambda$= $20 R^{-1}$
\cite{dienes}. Thus one often takes $\Lambda =20 R^{-1}$ as the
cut-off of the model. If one does not demand such unification, one can
extend the value of $\Lambda$ to about $40 R^{-1}$, above which the
$U(1)$ coupling becomes nonperturbative.

\subsection{{ Gravity} in extra dimensions}
In this section, we will consider the scenario where { gravity} is assumed to propagate in $N$ "large" extra dimensions compactiﬁed on a torus with volume $V_N=r^N$, where $r$ is the size of the extra dimensions\footnote{We have assumed that the extra dimensions have common size (symmetric torus), denoted by $r(\sim~{\rm few~eV}^{-1}~{\rm to~few~keV}^{-1})$. The generalization to an asymmetric torus with different radii is straightforward.}. The $(4+N)$-dimensional metric, assumed to be approximately flat, as $\hat g_{\hmu\hnu}=\eta_{\hmu\hnu}+\hk \hh_{\hmu\hnu}$, where $\hk^2=16\pi G^{(4+N)}$ and $G^{(4+N)}$ is the Newton constant in $(4+N)$ dimension. The $(4+N)$-dimensional tensor $\hh_{\hmu\hnu}$ consists of three parts: a $4$-dimensional tensor (the graviton  $h_{\mu\nu}$), $N$ vectors (the graviphotons $A_{\mu i}$) and $N^2$ scalars (the graviscalar $\phi_{ij}$):
\beq \hh_{\hmu\hnu}\ =\ V_N^{-1/2}\left(\begin{array}{cc}
h_{\mu\nu}+\eta_{\mu\nu}\phi & A_{\mu i}\\
A_{\nu j}   &  2 \phi_{ij}
\end{array}\right)\ ,
\label{hh_def}
\eeq
 where, $\mu, \nu=0,1,2,3$ and $i,j = 4,5,6,\ldots 3+N$ and $\phi = \phi_{ii}$.
In the similar way as discussed in the previous section, these fields are compactified on an $N$-dimensional torus $T^N$ and have the following KK-expansions:
\begin{eqnarray}
&& h_{\mu\nu}(x,y)\ =\ \sum_{\vec n} h_{\mu\nu}^{\vec{n}}(x)\ 
\exp\left(i {2\pi \vec{n}\cdot\vec{y}\over r}\right)\ ,\\
&& A_{\mu i}(x,y)\ =\ \sum_{\vec n} A_{\mu i}^{\vec{n}}(x)\ 
\exp\left(i {2\pi \vec{n}\cdot\vec{y}\over r}\right)\ ,\\ 
&& \phi_{ij}(x,y)\ =\ \sum_{\vec n} \phi_{ij}^{\vec{n}}(x)\ 
\exp\left(i {2\pi \vec{n}\cdot\vec{y}\over r}\right)\ , \qquad
{\vec n} = \{n_1,n_2,\cdots,n_N\}\ ,\label{mode}
\end{eqnarray}
where, the modes of $\vec{n}\neq 0$ are 
the KK-states\footnote{It is important to note that the kinetic terms of the gravity KK-excitations ($G^{\vec n}$: graviton, graviphotons and graviscalars) in the effective theory do not have their canonical form when the gravity Lagrangian is written
in terms of the fields $h_{\mu\nu}^\vn, A_{\mu i}^\vn$ and
$\phi_{ij}^\vn$. It is then necessary to redefine the fields in
the gravity sector and to work in terms of "physical" fields that have
canonical kinetic and mass terms. The details of this redefinition
are worked out in Ref.~\cite{HLZ,GRW}.} and the zero modes, $\vn=\vec{0}$,  correspond to the massless graviton, graviphotons and graviscalars in $4$D effective theory. The extra dimensional derivatives in the kinetic terms of the gravity Lagrangian results into a mass $m_{\vec n}=2\pi |\vec n|/r$ for the level-$\vec n$ graviton, graviphotons and graviscalars. After discussing the KK-expansion of the matter and gravity fields, we are now equiped enough to compute the interactions of gravity with matter.

\subsection{Gravity-matter interactions}
There are three possible ways of embedding matter fields in the extra dimensions (the bulk). 
\begin{itemize}
\item In the usual ADD scenario, the SM matter fields (fermions and
bosons) are restricted to a $4$D subspace (3-brane), called the SM brane, embedded in the $(4+N)$D bulk. For the case of matter restricted on the 4D SM brane, the interaction Lagrangian has been computed in Refs. (\cite{HLZ,GRW}).
\item One could naturally construct a more general theory by allowing the SM fields to also propagate in the whole space (the bulk). This would imply that the SM particles also acquire a KK-tower of excitations. However, one does not observe such excitations in colliders. This implies that either the SM
fields do not propagate in the bulk or the scale on which they propagate is
much smaller ($\sim$ TeV$^{-1}$) than the scale associated with gravity, so that the masses of KK-excitations of matter are high enough to evade the experimental constraints. 
\item In the { fat brane scenario}, there are $N$ extra dimensions of eV$^{-1}$ size, into which gravity propagates. However, matter propagation is restricted only to a small length ($\sim$ TeV$^{-1}$), associated with the thickness of the brane along these extra dimensions. For this scenario, the gravity-matter interaction Lagrangian is derived in Refs. (\cite{GUED1,GUED2}). In the following, we will briefly discuss this scenario.  
\end{itemize}

Let us assume that there is only one small extra dimension, denoted by $y=x^4$, of size $\pi R\sim$~TeV$^{-1}$, in which both matter and gravity propagates and that there are $N$ larger extra dimensions, denoted by $x^5,x^6,\ldots,x^{4+N}$, of size $r\sim$~eV$^{-1}$, in which only gravity propagates. The phenomenology of this scenario is governed by three parameters, namely the number of extra dimension ($N$) and the sizes of the small and large extra dimension: $R$ and $r$, respectively. The size of large extra dimension ($r$) is related the $4$D and $(4+N)$D Planck scale ($M_{Pl}$ and $M_D$, respectively) by the ADD relation:
\begin{equation}
M_{Pl}^2~=~M_D^{N+2}\left(\frac{r}{2\pi}\right)^N.
\label{ADDrelation}
\end{equation}
In this scenario, the gravitational coupling of the matter fields (fermions and bosons) is given by,
\begin{equation}
S_{int}=\int d^{4+N}x \ \delta(x^5) \ldots \delta(x^{4+N})\ \sqrt{-\hat{g}}\ {\cal L}_m\ ,
\label{coup}
\end{equation}
where $\hat{g}$ is the metric in $(4+N)$D and ${\cal L}_m$ is the matter Lagrangian. The ${\cal O}(\hk)$ term of Eq.~\ref{coup} is given by,
\beq
{\cal S}_{int} \supset -{{\hat \kappa} \over 2}
\int d^{4+N} x \ \delta(x^5) \ldots \delta(x^{4+N})\
 \hh^{\hmu\hnu} T_{\hmu\hnu} \ ,
\label{act_gen}
\eeq
where, $T_{\hmu\hnu}$ is the { energy-momentum (EM)} tensor in $(4+N)$D and defined as,
\beq
T_{\hmu\hnu} \ = \ \left( -\hat{\eta}_{\hmu \hnu} + 2 \frac{\del 
{\cal L}_m}{\del \hat{g}^{\hmu \hnu} }\right)_{\hat{g} = \hat{\eta}} \ .
\label{em_tensor}
\eeq
Since, matter propagation is restricted to only one { extra-dimension} ($x^4$), the interaction action can be written in terms of $(\mu \nu),\ (\mu 4)$ and $(44)$ components of the matter { EM} tensor:
\beq
 {\cal S}_{int} \supset -{\kappa \over 2} \int d^4 x
\int_0^{\pi R} dy  \sum_{\vn} \left[ \left( h^\vn_{ \mu \nu} +
\eta_{\mu \nu} \phi^\vn \right) T^{\mu \nu} -2 A^\vn_{\mu 4}
T^\mu_4 + 2 \phi^\vn_{44} T_{44} \right] e^ {2\pi i { n_4 y\over
r}}.
\label{int_action}
\eeq
Here $\kappa$ is the four-dimensional Newton's constant 
$\kappa^2 \ \equiv \ {16 \pi G^{(4)}} \ = \ V_n^{-1/2} \hat{\kappa}$.

For a given matter Lagrangian, ${\cal L}_m$, which is the { mUED} Lagrangian in our case, it is straightforward to obtain the matter { EM} tensor using Eq.~\ref{em_tensor}. Expanding the matter fields in KK-modes and integrating over the coordinate $y~(x^4)$, one can work out the { Gravity-matter} Feynman interaction rules. The resulting Feynman interaction rules are quite complicated and can be found in Ref.~\cite{GUED1}.

\section{Phenomenology}
After introducing the model, we have all the necessary ingredients to discuss the phenomenology of this model. In this section, we will concentrate only on the phenomenology of the level-1 matter fields. In the preceding section, we identified the SM doublet and singlet quarks with the level-0 excitation of the 5D fields $Q$, $u$ and $d$ respectively (see Appendix~\ref{correction}). Similar would be the case for the leptonic fields. The level-1 fermionic sector thus constitutes of $Q_{1}$, $u_{1}$, $d_{1}$, $L_{1}$ and $e_{1}$. For the corresponding bosonic sector, we have, the level-1 Higgs and gauge bosons (excited gluon: $g_1$, W-bosons: $W^{\pm}_1~{\rm and~}Z_1$ and photon: $\gamma_1$)
excitations. 
\begin{figure}[t]
\begin{center}
\epsfig{file=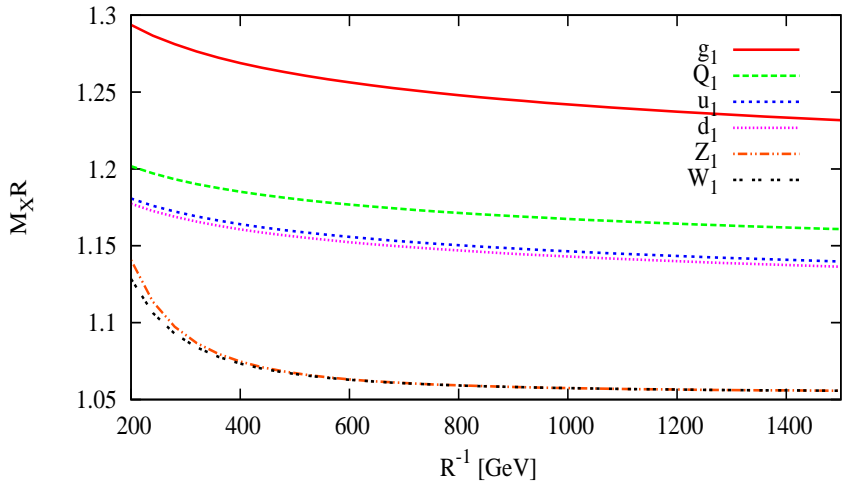,width=7cm,height=12cm,angle=270}
\end{center}
\caption{Variation of $M_XR$ (where $X$ corresponds to either
  $g_1,~Q_1,~u_1$, $d_1$, $W_1$ or
  $Z_1$) as a function of $R^{-1}$ for $\Lambda \,R=20$. Here, $Q_1,~{\rm and}~u_1$ does not include the top's partners.}
\label{fig:mass}
\end{figure}

While the tree-level masses, in the absence of electroweak 
symmetry breaking, would be $R^{-1}$ for each, the inclusion of radiative 
corrections does change them~\cite{Cheng:2002iz}. The KK-fermions receive mass corrections from the gauge interactions (with KK-gauge bosons) and Yukawa interactions. All of these give positive mass shift. The gauge fields receive mass corrections from the self-interactions and gauge interactions (with KK-fermions). Gauge interactions with fermions give a negative mass shift, while the self-interactions give positive mass shift (see Appendix \ref{correction}). However, masses of the hypercharge gauge boson $\gamma_1$ receive only negative corrections from fermionic loops. Numerical computation shows that
the lightest KK-particle is the hypercharge gauge boson $\gamma_1$.
The radiative corrections are dependent on the cutoff scale $\Lambda$ 
(note that an ultraviolet completion needs 
to be defined for all such theories). We present  
the corrections for $\Lambda=20R^{-1}$. To be specific, 
\begin{equation}  
M_{L_{1}}\simeq{1.03}~{R^{-1}}, \hskip 30pt 
M_{e_{1}}\simeq{1.01}~{R^{-1}},\hskip 30pt
M_{\gamma_{1}}~\simeq~{1.00}~{R^{-1}}, 
\label{mass}
\end{equation}
with the numerical factors being almost independent of $R^{-1}$.   
For the other colored states, an additional mild dependence accrues
from the scale dependence of the QCD coupling constant. 
For the $SU(2)$ gauge bosons, the {$R^{-1}$}
dependence arises from the non-zero mass of the SM $W^{\pm}$ and $Z$-boson. 
In Fig.~\ref{fig:mass}, we present these masses as a
function of $R^{-1}$ with {$\alpha_s = \alpha_s (M_X)$.}

\subsection{Decay of level-1 particles}
In the framework of the present model, the decay of the KK-particles can be classified into two different categories: {\bf Category 1:} { KK-number Conserving Decay (KKCD)} and {\bf Category 2:} { Gravity Mediated Decay (GMD)}. In the following, we will discuss these two categories in some details.

\subsubsection{{\bf Category 1:} {\em KK-number Conserving Decay (KKCD)}}
Conservation of { KK-number} (as well as { KK-parity}) allows level-1 particles to decay only into a lighter level-1  particle and one or more SM particles if kinematically allowed. The { KK-number Conserving Decays} of  level-1 particles have been investigated in detail in Ref.~\cite{KKCD}. It is clear from Eq. \ref{mass} that $\gamma_1$ is the lightest KK-particle (LKP) in this theory. Therefore, the KK-number Conserving Decays of all the level-1 particles result in one or more SM particles plus $\gamma_1$. In the following, we will briefly discuss the decays of the different level-1 KK-particles.

Typical mUED spectrum shows that the colored KK-states are heavier than the electroweak KK-particles and level-1 gluon $g_1$ is the heaviest (see Eq.~\ref{mass} and Fig.~\ref{fig:mass}). It can decay to both singlet ($u_1$ and $d_1$) and doublet ($Q_1$) quarks with almost same probability, although there is a slight kinematic preference to the singlet channel. The singlet quark can decay only to $\gamma_1$ and SM quark. On the other hand, doublet quarks decay mostly to the KK-excitation of electroweak gauge bosons, namely the $W_1$ or $Z_1$. Hadronic decay modes of $W_1$ and $Z_1$ are closed kinematically (see Fig.~\ref{fig:mass}) and these can decay universally to all doublet lepton flavors ($L_1$). 
The KK-leptons finally decay to $\gamma_1$ and a ordinary (SM) lepton. Since $\gamma_1$ is the LKP, further { KK-number Conserving Decay} of $\gamma_1$ is forbidden. 

\begin{figure}[t]
\begin{center}
\epsfig{file=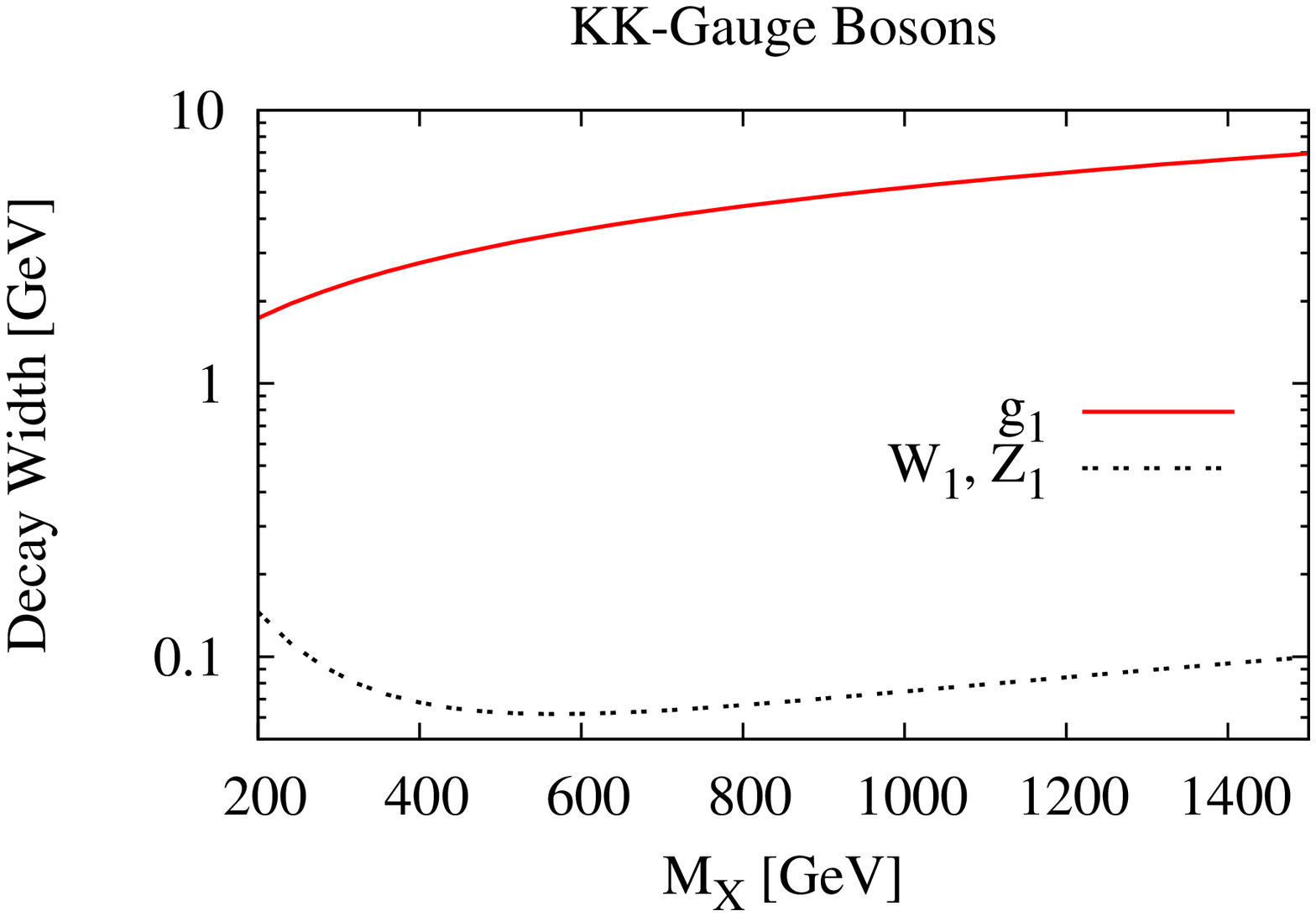,width=5cm,height=5.2cm,angle=270}
\epsfig{file=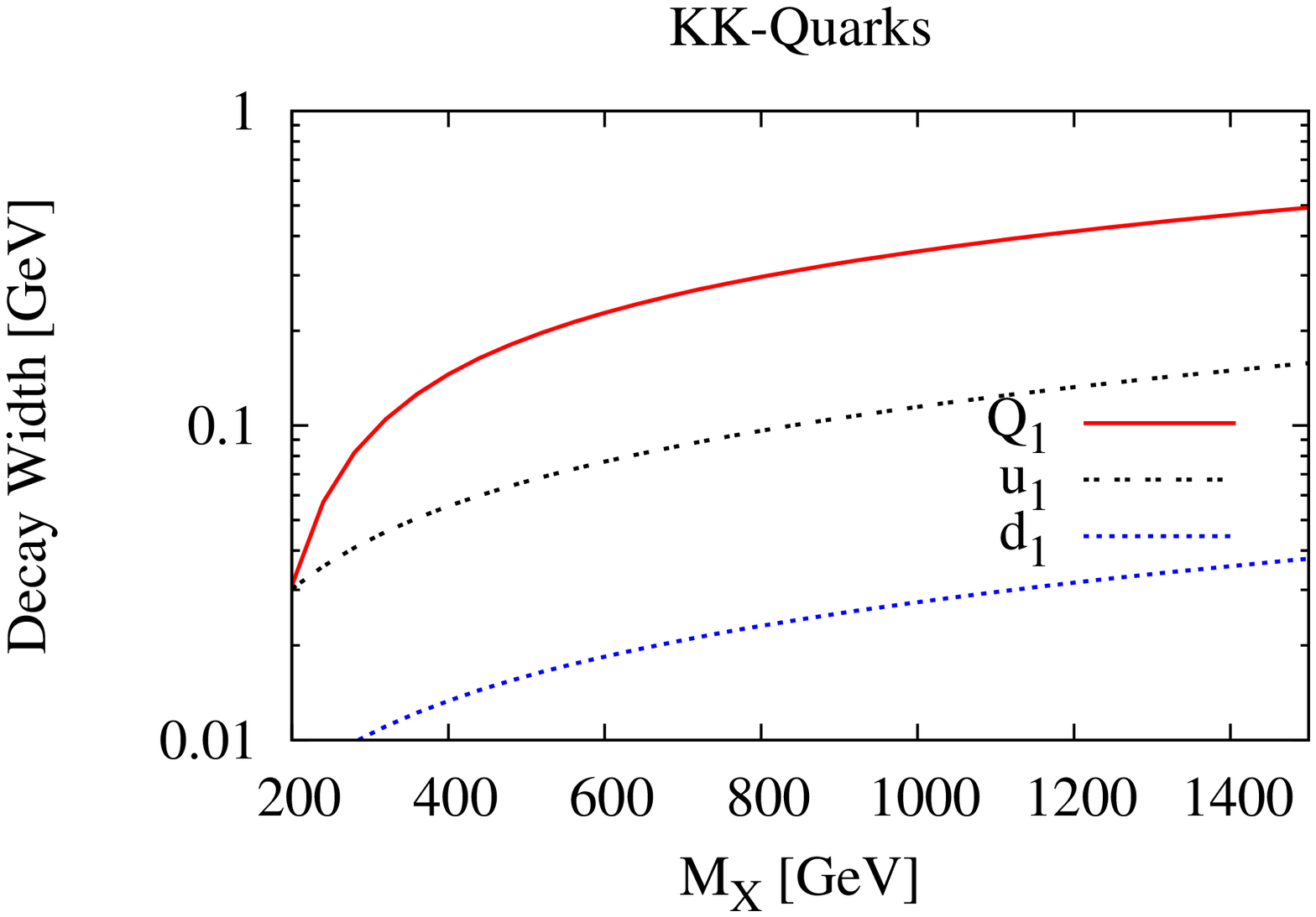,width=5cm,height=5.2cm,angle=270}
\epsfig{file=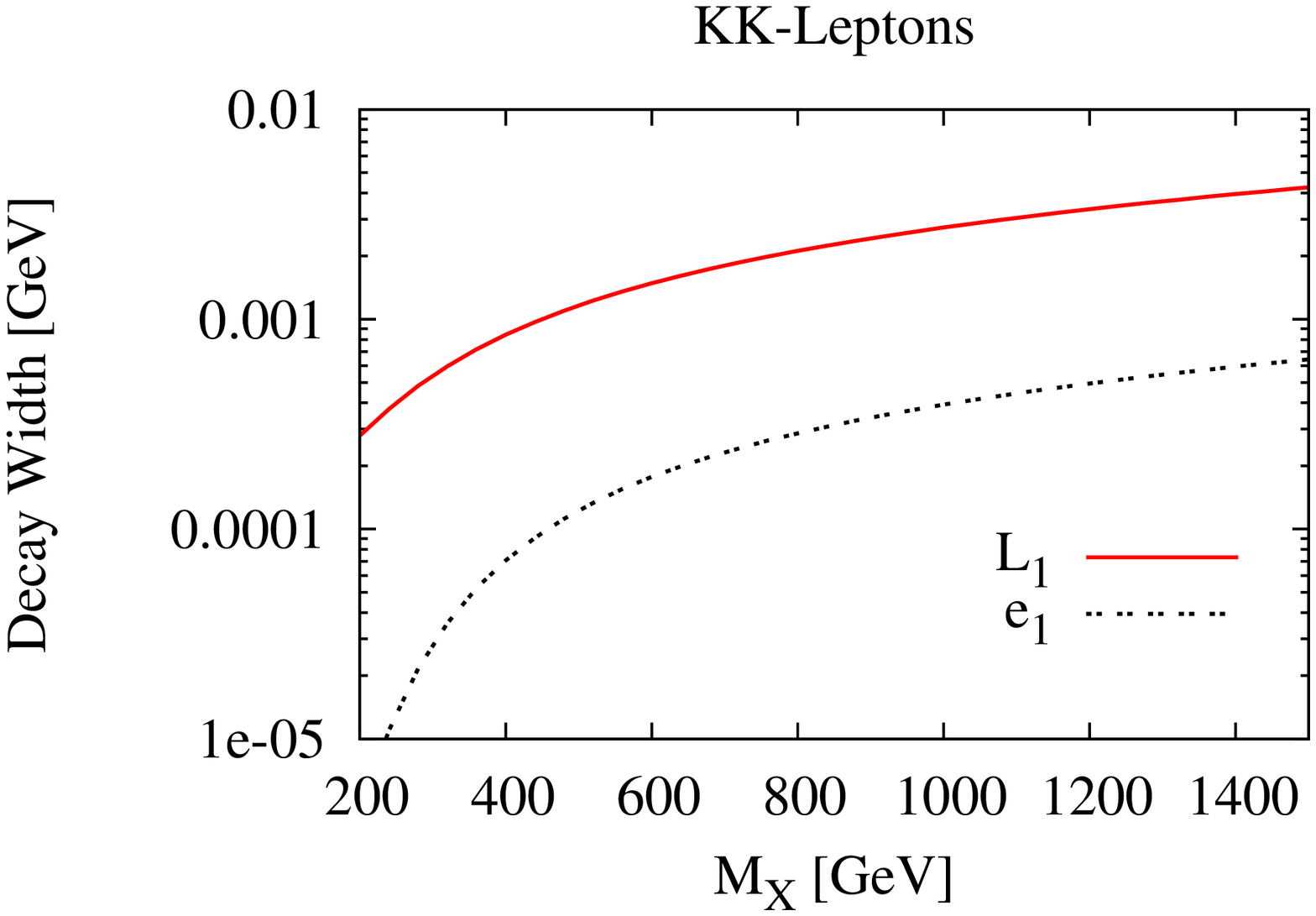,width=5cm,height=5.2cm,angle=270}
\end{center}
\caption{{ KK-Number Conserving Decay (KKCD)} widths for (left panel) KK-bosons, (middle panel) KK-quarks and (right panel) KK-leptons as a function of the KK-particle mass.}
\label{fig:KKCwidth}
\end{figure} 

In Fig.~\ref{fig:KKCwidth}, we present KK-number conserving total decay widths for (left panel) KK-bosons, (middle panel) KK-quarks and (right panel) KK-leptons as a function of the KK-particle mass. The important features of different decay widths are summarized below.
\begin{itemize}
\item Since the decay of $g_1$ is purely a QCD driven process, the total KKCD width of $g_1$ is a few orders of magnitude higher than the decay widths of other KK-particles.  
\item Level-1 $SU(2)$ gauge bosons ($W_1$ and $Z_1$) decay in to level-1 doublet leptons ($L_1$). Since the mass splitting between $W_1~(Z_1)$ and $L_1$ is very small (see Eq.~\ref{mass} and Fig.~\ref{fig:mass}), this decay width is kinematically suppressed for the intermediate $R^{-1}$ ($400<R^{-1}<800$). For small $R^{-1}$, $W_1~(Z_1)$ mass gets significant contribution from the SM $W~(Z)$-boson mass (see Fig.~\ref{fig:mass}) and therefore, the approximate degeneracy between $W_1~(Z_1)$ and $L_1$ mass is partially removed in the low $R^{-1}$ region. This can be attributed to the fact that this decay width increases as we lower the value of $R^{-1}$ (see Fig.~\ref{fig:KKCwidth} left panel). 
\item Doublet quarks ($Q_1$) decay into $W_1$ or $Z_1$. Fig.~\ref{fig:KKCwidth} (middle panel) shows that this decay widths are kinematically suppressed for low $R^{-1}$.
\item Singlet quarks ($u_1$ and $d_1$) decay into $\gamma_1$. These decay width is proportional to the square of the singlet quark hypercharge. As a result, the total decay width of $u_1$  is larger than the decay width of $d_1$ by a factor of $4$. 
\end{itemize}

\begin{figure}[t]
\begin{center}
\epsfig{file=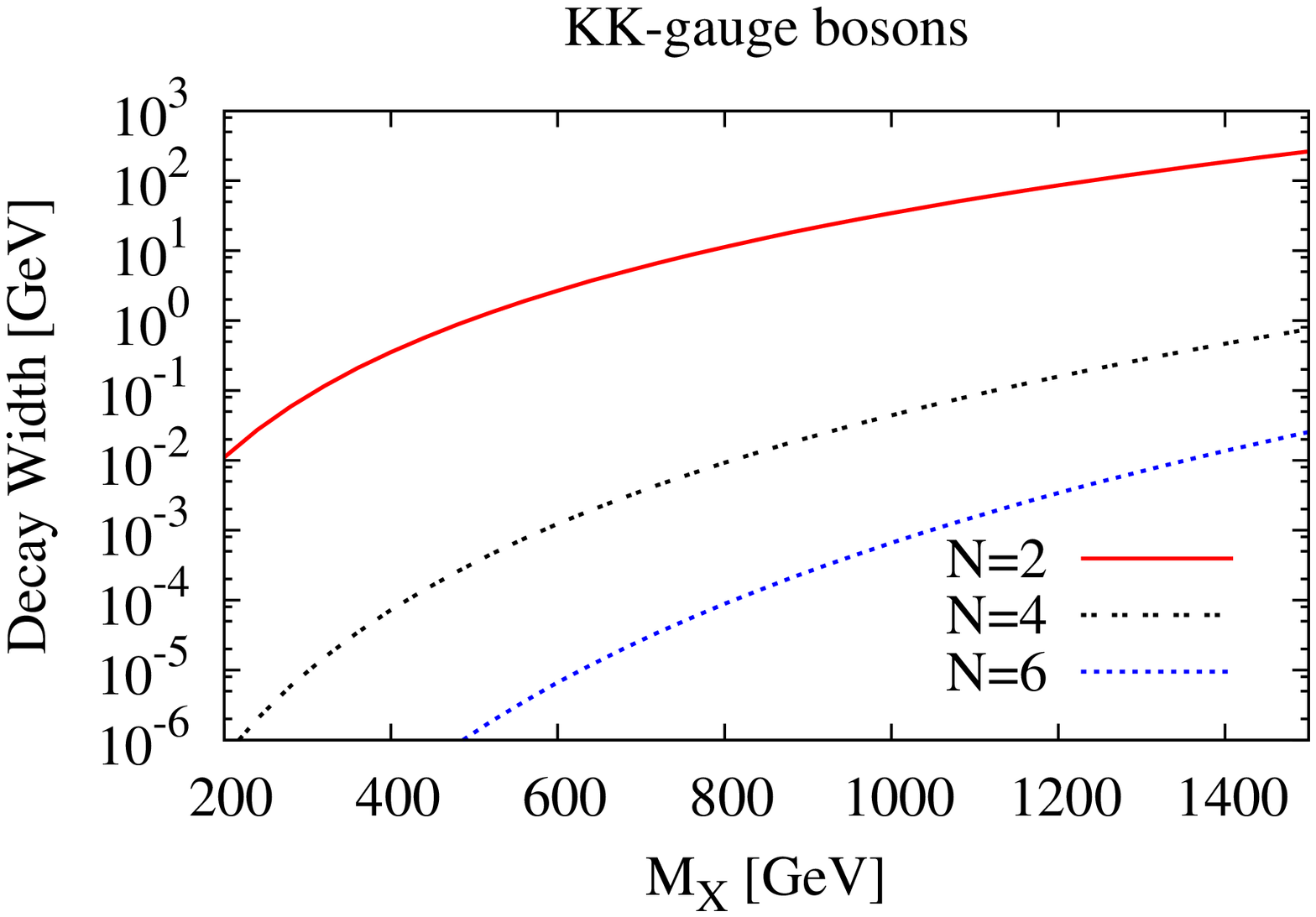,width=7cm,height=7cm,angle=270}
\epsfig{file=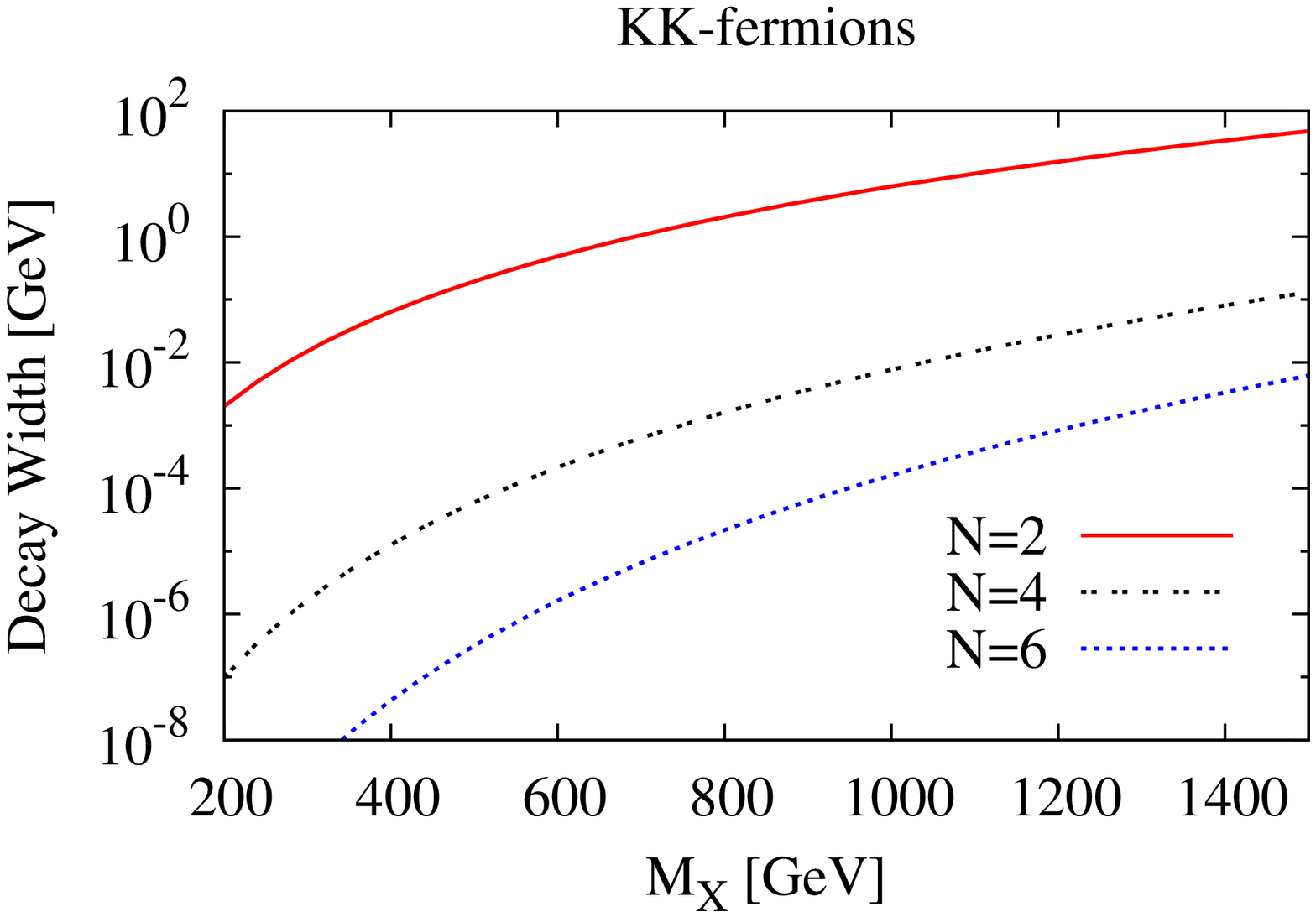,width=7cm,height=7cm,angle=270}
\end{center}
\caption{{ Gravity Mediated Decay (GMD)} widths for (left panel) KK-gauge bosons and (right panel) KK-fermions for three different number ($N=2,~4~{\rm and}~6$) of { "large"} extra dimension as a function of the KK-particle mass. In this plot, we have assumed the fundamental ($4+N$)D Planck mass $M_D=5$ TeV.}
\label{fig:GMDwidth}
\end{figure} 

\subsubsection{{\bf Category 2:} {\em Gravity Mediated decay (GMD)}}

In the framework of the model described in section 2, the universal extra dimension is assumed to be the thickness of a 3-brane (in which the SM particles propagate) embedded in a $(4+N)$D (in which gravity propagates). The specific positioning of the 3-brane in the extra dimensions is a breakdown of translational symmetry along the extra dimension accessible to the SM fields. Therefore, { KK-number} is no more a conserved quantity for the { gravity-matter} interactions. This has extremely interesting phenomenological consequences. Gravity interactions will mediate the decay of the level-1 KK-excitations of matter into graviton excitation and the SM particles.

The partial decay width of level-1 matter field (fermion, gauge boson or scalar) into a level-$\vec n$ gravity excitation $G^{\vec n}$ ($G^{\vec n}\ni$ graviton, graviphoton or graviscalar) and respective SM matter field can be computed using the Feynman rules for the gravity-matter interactions derived from the { action} in Eq.~\ref{int_action}. The total decay width is obtained by summing over all possible gravity excitations with mass smaller that the decaying particle: 
\beq
\Gamma = \sum_{\vec n} \Gamma_{\vec n}=\sum_{\vec n} \left [\Gamma_{h^{\vec n}}+\Gamma_{A^{\vec n}}+\Gamma_{\phi^{\vec n}}\right],
\eeq
where, $\Gamma$ is the total gravity mediated decay width and $\Gamma_{h^{\vec n}}$, $\Gamma_{A^{\vec n}}$ and $\Gamma_{\phi^{\vec n}}$ are the partial decay widths into level-$\vec n$ graviton, graviphoton and graviscalar respectively. The gravity KK-states are nearly degenerate in mass and the mass splitting is given by $\Delta m = 2\pi/r\sim$ eV to keV. Therefore, the sum can be replaced by an integral:
\beq
\sum_{\vec n} \Gamma_{\vec n} \longrightarrow \int \Gamma_{\vec n}~ d^N \vec n,
\eeq
where, $d^N \vec n$ is the number of gravity excitations with masses in a range
($m_{\vec n},~ m_{\vec n} + dm$). The mass of level-$\vec n$ gravity excitation is given by, $m_{\vec n}^2=4\pi^2 \vec n^2/r^2$ and therefore, $\vec n^2=m_{\vec n}^2/\Delta m^2$. The number of gravity excitations with masses in a range
($m_{\vec n},~ m_{\vec n} + dm$) is given by the volume of the annular region between two $N$-dimensional hyper-sphere of radius $m_{\vec n}/\Delta m~{\rm and}~ (m_{\vec n} + dm)/\Delta m$:
\beq
d^N \vec n~=~\left(\frac{m_{\vec n}}{\Delta m}\right)^{N-1}~\frac{dm}{\Delta m}~d\Omega~=~\frac{1}{\Delta m^N}~m_{\vec n}^{N-1}~ dm~d\Omega,
\eeq
where, $d\Omega$ is the $N$-dimensional angular element. Using Eq.~\ref{ADDrelation}, we obtain $\Delta m^N=M_D^{N+2}/M_{Pl}^2$. Therefore, the total gravity mediated decay width is given by,
\beq
\Gamma=\frac{M_{Pl}^2}{M_D^{N+2}}~\int ~\Gamma_{\vec n}~m_{\vec n}^{N-1}~dm~d\Omega.
\label{totalwidth}
\eeq

The gravity mediated decays of the KK-particles were previously computed in Ref~\cite{GUED1,GUED2}. For computing the gravity mediated decay widths of the level-1 KK-particles, we have followed the analysis of Ref.~\cite{GUED1,GUED2}. The expressions for the gravity mediated partial decay widths of level-1 fermions and gauge bosons can be found in Ref.~\cite{GUED2}. In Fig.~\ref{fig:GMDwidth}, we have presented the gravity mediated decay widths of the level-1 (left panel) KK-gauge bosons and (right panel) KK-fermions as a function of KK-particle mass. We have considered three different number ($N=2,~4~{\rm and}~6$) of { "large"} extra dimension accessible for the { gravity}. For computing the gravity mediated decay widths of Fig.~\ref{fig:GMDwidth}, we took the fundamental ($4+N$)D Planck mass, $M_D=5$ TeV. Fig.~\ref{fig:GMDwidth} clearly shows that the gravity mediated decay widths are larger for $N=2$ in the case of both KK-fermions and gauge bosons. This can be attributed to the fact that for the lower values of $N$, the splittings between the gravity excitations masses are smaller and hence, the density of gravity KK-states are larger. Even for $N = 6$, the decay widths are typically large enough that the particles will decay in the detector. The important features of the gravity mediated decay widths of KK-gauge bosons and fermions are summarized below.
\begin{itemize}
\item The { gravity-matter} interaction action in Eq.~\ref{int_action} includes coupling between KK-gluon, SM-gluon and a level-$\vec n$ gravity excitation $G^{\vec n}$. Therefore, level-1 KK-gluon can decay into a SM-gluon and a level-$\vec n$ gravity excitation. The total decay widths are presented in Fig.~\ref{fig:GMDwidth} (left panel) for $N=2,~4,~{\rm and}~6$.
\item Similarly, level-1 $W$-boson ($W_1^\pm$) decays in to a SM $W^\pm$ and gravity excitations. Numerically, this decay width is almost equal to the decay widths presented in Fig.~\ref{fig:GMDwidth} (left panel). Very small deviation arises due to the non-zero mass of the SM $W^\pm$-boson. 
\item It has been already discussed in details in section 2.1 that $Z_1$ and $\gamma_1$ are almost purely the level-1 excitations of $W^{3}_\mu$ and $B_\mu$. Therefore, both $Z_1$ and $\gamma_1$ can decay into a photon or $Z$-boson in association with a level-$\vec n$ gravity excitation. The decay widths presented in Fig.~\ref{fig:GMDwidth} (left panel) also correspond to the total { gravity mediated decay} widths for $Z_1$ and $\gamma_1$.
\item The KK-fermions (quarks as well as leptons) decay into the corresponding SM fermions and gravity excitations. The total { gravity mediated decay} widths (for $N=2,~4~{\rm and}~6$) for the  level-1 KK-fermions are presented in Fig.~\ref{fig:GMDwidth} (right panel).
\end{itemize}

\subsection{Collider Signature}

After discussing the decays of the level-1 KK-particles, we are now equipped enough to discuss the collider signature of this scenario. The KK-quarks and gluons carry colors and it is needless to mention that their production cross sections are high at the LHC. Therefore, at the LHC, the dominant production processes of { mUED} are the pair production of level-1 KK-quarks and KK-gluons. The tree level KK-number conserving couplings of KK-quarks and gluons are similar to the SM couplings and there is no $\Lambda$ or $R$ dependence in their couplings. However, the masses of the KK-states are logarithmically $\Lambda$ dependent (see Appendix~\ref{correction}) and thus, the KK-production cross sections depend mildly on the cut-off $\Lambda$. In this analysis, we have considered $\Lambda=20 R^{-1}$.

\begin{figure}
\begin{center}
\epsfig{file=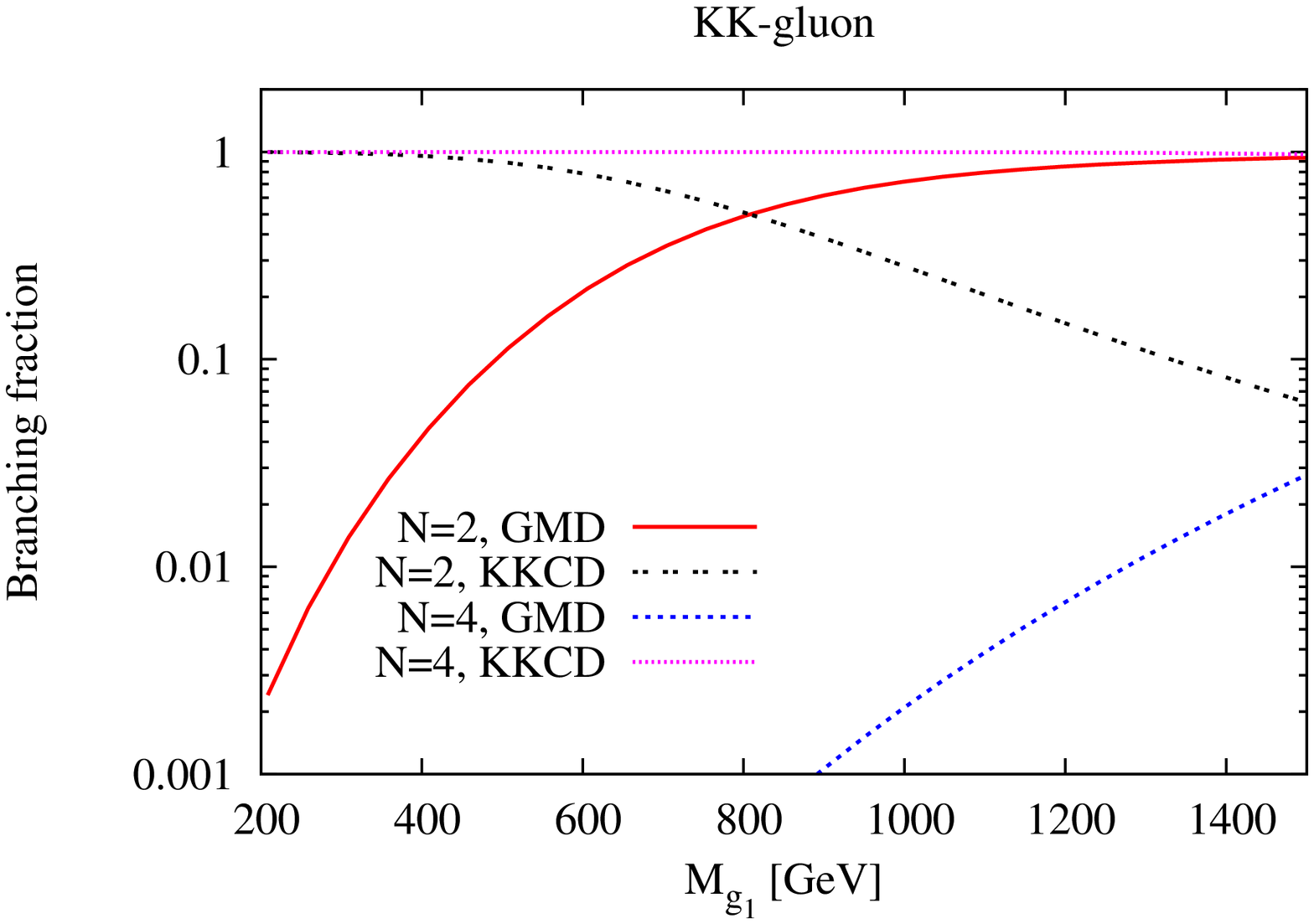,width=7cm,height=7cm,angle=270}
\epsfig{file=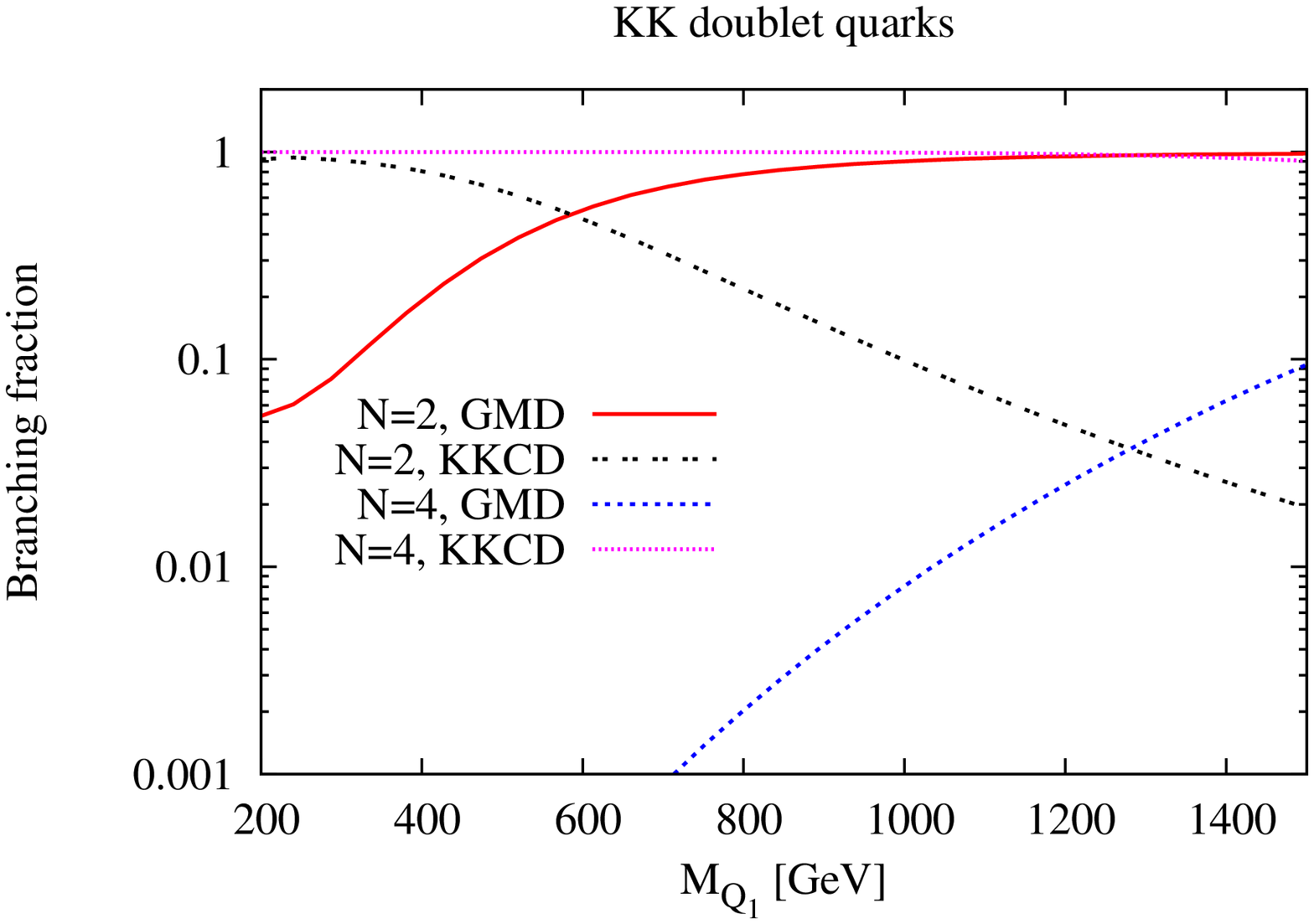,width=7cm,height=7cm,angle=270}\\
\epsfig{file=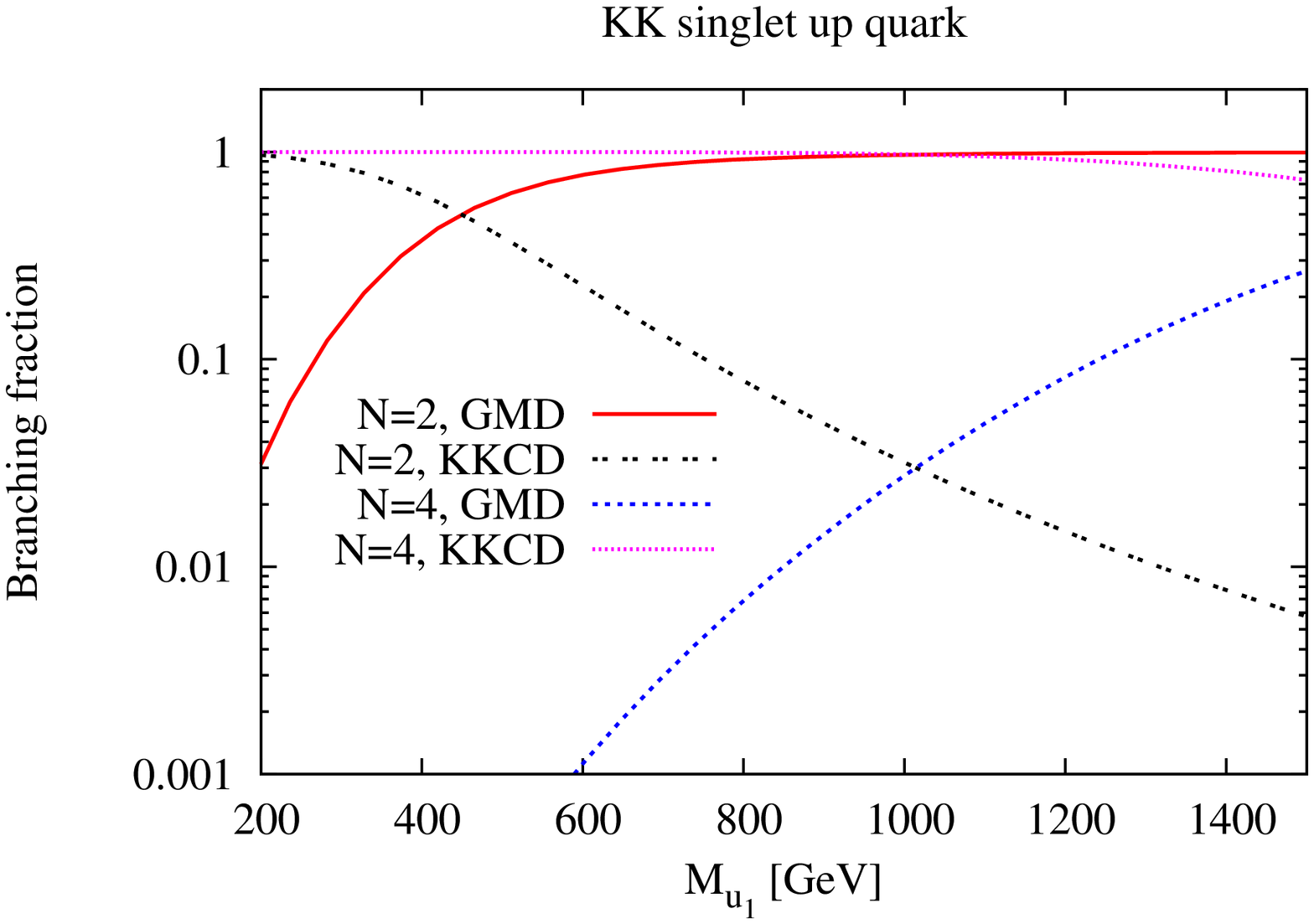,width=7cm,height=7cm,angle=270}
\epsfig{file=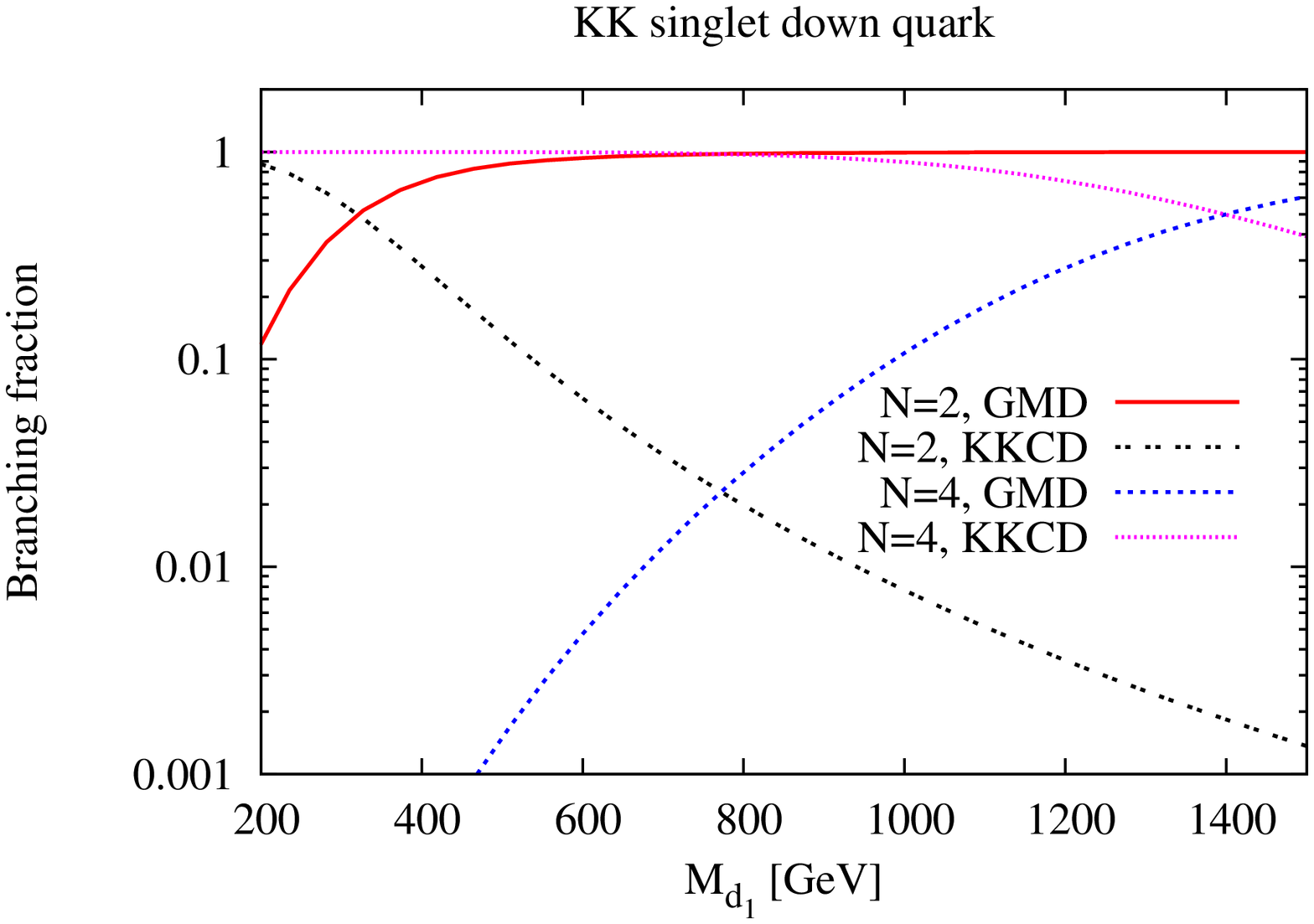,width=7cm,height=7cm,angle=270}
\caption{{ Gravity Mediated Decay (GMD)} branching fractions and { KK-Number Conserving Decay (KKCD)} branching fractions for colored level-1 KK-particles ($g_1$: Top Left, $Q_1$: Top Right, $u_1$: Bottom Left and $d_1$: Bottom Right) for $N=2~{\rm and}~4$ as a function of KK-particle mass. In this plot, we have assumed the fundamental ($4+N$)D Planck mass $M_D=5$ TeV.}
\label{Branching}
\end{center}
\end{figure}

After being produced, the level-1 KK-quarks and gluons decay into lighter KK-particles (level-1 KK-matter fields or level-$\vec n$ KK-gravity excitations) in association with one or more SM particles. The decays of the level-1 KK-particles was discussed in details in the previous section. Comparison of { KK-number Conserving Decay} widths (see Fig.~\ref{fig:KKCwidth}) and { Gravity Mediated Decay} widths (see Fig.~\ref{fig:GMDwidth}) of the level-1 KK-matter fields shows that for $N=6$, the { GMD} widths are a few orders of magnitude smaller than the { KKCD} widths. However, for $N=2~{\rm and}~4$, the { GMD} widths are comparable with the { KKCD} widths. Therefore, in the context of a collider experiment, depending on the number of { "large"} extra dimensions, $N$, different final state signal topologies are possible.
\begin{itemize}
\item {\bf No. of { "large"} extra dimensions $N=6$:} In this case, the { gravity mediated decay} widths are suppressed compared to { KK-number conserving decay} widths. Therefore, level-1 matter fields decay cascades involving other level-1 particles until reaching the LKP ($\gamma_1$) at the end of the decay chain. In the presence of gravity induced interactions, $\gamma_1$ further decays into a level-$\vec n$ gravity excitation ($G^{\vec n}$) in association with a photon (with branching fraction $\sim$ 78\%) or $Z$-boson (with branching fraction $\sim$ 22\%). The decay cascades of level-1 gluon ($g_1$) are schematically shown in Fig.~\ref{cascade}. With two decay chains per event, the final state would be characterized by a pair of photon or $Z$-boson ($\gamma \gamma,~ZZ~{\rm or}~\gamma Z$) in association with a few leptons, jets and large missing transverse momentum. The missing transverse momentum results from the escaping gravity excitations, whereas, the leptons and jets are emitted during the cascade decays.
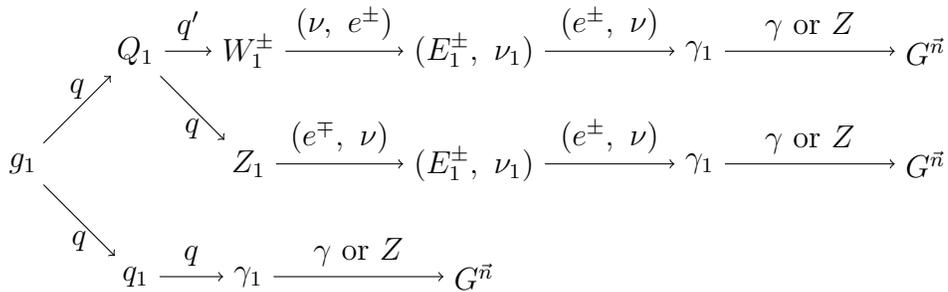
\begin{figure}[h]
 \begin{center}
  \begin{tikzpicture}[node distance=1.5cm, auto]
\node (a1){$g_1$};
\node (a2)[right of=a1]  {};
\node (a3)[above of=a2]  {$Q_1$};
\node (a4)[below of=a2]  {$q_1$};
\node (a31)[right of=a3]  {$W^\pm_1$};
\node (b2)[below of=a3]  {};
\node (a32)[right of=b2]  {$Z_1$};
\node (b3)[right of=a31]  {};
\node (a33)[right of=b3]  {$(E_1^\pm,~\nu_1)$};
\node (b4)[right of=a32]  {};
\node (a34)[right of=b4]  {$(E_1^\pm,~\nu_1)$};
\node (b5)[right of=a33]  {};
\node (b6)[right of=a34]  {};
\node (a35)[right of=b5]  {$\gamma_1$};
\node (a36)[right of=b6]  {$\gamma_1$};
\node (a37)[right of=a4]  {$\gamma_1$};
\node (b8)[right of=a35]  {};
\node (b9)[right of=a36]  {};
\node (b10)[right of=a37]  {};
\node (a38)[right of=b8]  {$G^{\vec n}$};
\node (a39)[right of=b9]  {$G^{\vec n}$};
\node (a40)[right of=b10]  {$G^{\vec n}$};

\draw[->](a1) -- node[above]{$q$} (a3);
\draw[->](a1) -- node[below]{$q$}(a4);
\draw[->](a3) -- node[above]{$q^\prime$} (a31);
\draw[->](a3) -- node[below]{$q$} (a32);
\draw[->](a31) -- node[above]{$(\nu,~e^\pm)$} (a33);
\draw[->](a32) -- node[above]{$(e^\mp,~\nu)$} (a34);
\draw[->](a33) -- node[above]{$(e^\pm,~\nu)$} (a35);
\draw[->](a34) -- node[above]{$(e^\pm,~\nu)$} (a36);
\draw[->](a4) -- node[above]{$q$} (a37);
\draw[->](a35) -- node[above]{$\gamma~{\rm or}~Z$} (a38);
\draw[->](a36) -- node[above]{$\gamma~{\rm or}~Z$} (a39);
\draw[->](a37) -- node[above]{$\gamma~{\rm or}~Z$} (a40);

 \end{tikzpicture}
  \caption{Decay cascade of level-1 gluon ($g_1$) for $N=6$, where $G^{\vec n}\ni h^{\vec n},~A^{\vec n}~{\rm or}~\phi^{\vec n}$.}
  \label{cascade}
 \end{center}
\end{figure}

\item {\bf No. of { "large"} extra dimensions $N=2~{\rm or}~4$:} In this case, the { GMD} widths are comparable with the { KKCD} widths. In Fig.~\ref{Branching}, we have plotted the { GMD} and { KKCD} branching fractions for colored level-1 KK-particles ($g_1$: Top Left, $Q_1$: Top Right, $u_1$: Bottom Left and $d_1$: Bottom Right) as a function of KK-particle mass. We have consider both $N=2~{\rm and}~4$. Fig.~\ref{Branching} (see top panel) shows that for $N=2$, { GMD} modes for $g_1~(Q_1)$ dominate over the { KKCD} modes for $M_{g_1(Q_1)} > 800~(600)$ GeV. Whereas, for the singlet KK-quarks $u_1$ and $d_1$ (see Fig.~\ref{Branching} bottom panel), { GMD} modes become dominant for singlet KK-quark mass above $500$ and $300$ GeV respectively. Fig.~\ref{Branching} also shows that for $N=4$, { GMD} modes become significant only for the large values of the KK-particle mass. Therefore, for $N=2~{\rm or}~4$, the decay cascade of level-1 KK-particles involves both { GMD} and { KKCD} modes at each step. Whereas, for $N=6$, the { GMD} modes appears only at the last step of the decay chain. In Fig.~\ref{cascade_KKCD}, we schematically present the decay cascade of $g_1$ for $N=2~{\rm or}~4$. Fig.~\ref{cascade_KKCD} suggests that in addition to the $\gamma\gamma~(ZZ,~\gamma Z)+p_T\!\!\!\!\!\!/~~$ signature (discussed in the previous paragraph), different other interesting signals are possible for $N=2$ and $4$. As for example, di-jet$+p_T\!\!\!\!\!\!/~~$ signature results when both produced level-1 particles (KK-gluon or quarks) directly decays into gravity excitations. On the other hand, jets$+\gamma (Z)+p_T\!\!\!\!\!\!/~~$ signal arises when one of the produced KK-particle follows KK-number conserving decay chain and other decays via gravity induced interactions.   
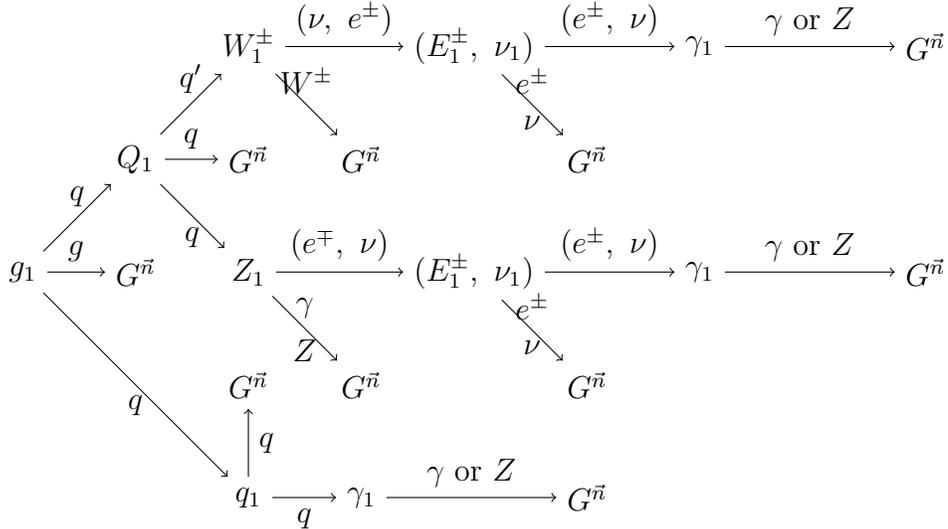
\begin{figure}[h]
 \begin{center}
  \begin{tikzpicture}[node distance=1.5cm, auto]
\node (a1){$g_1$};
\node (c1)[right of=a1]  {};
\node (a2)[right of=a1]  {};
\node (c2)[right of=c1]  {};
\node (c3)[below of=c2]  {};
\node (a3)[above of=a2]  {$Q_1$};
\node (a4)[below of=c3]  {$q_1$};
\node (b1)[right of=a3]  {};
\node (a31)[above of=b1]  {$W^\pm_1$};
\node (b2)[below of=a3]  {};
\node (a32)[right of=b2]  {$Z_1$};
\node (b3)[right of=a31]  {};
\node (a33)[right of=b3]  {$(E_1^\pm,~\nu_1)$};
\node (b4)[right of=a32]  {};
\node (a34)[right of=b4]  {$(E_1^\pm,~\nu_1)$};
\node (b5)[right of=a33]  {};
\node (b6)[right of=a34]  {};
\node (a35)[right of=b5]  {$\gamma_1$};
\node (a36)[right of=b6]  {$\gamma_1$};
\node (a37)[right of=a4]  {$\gamma_1$};
\node (b8)[right of=a35]  {};
\node (b9)[right of=a36]  {};
\node (b10)[right of=a37]  {};
\node (a38)[right of=b8]  {$G^{\vec n}$};
\node (a39)[right of=b9]  {$G^{\vec n}$};
\node (a40)[right of=b10]  {$G^{\vec n}$};

\node (d1) [right of=a1] {$G^{\vec n}$};
\draw[->](a1) -- node[above]{$g$} (d1);
\node (e3) [above of=a4]{};
\node (d2) [right of=a3] {$G^{\vec n}$};
\node (d3) [above of=a4] {$G^{\vec n}$};
\draw[->](a3) -- node[above]{$q$} (d2);
\draw[->](a4) -- node[right]{$q$} (d3);
\node (e4) [below of=a31]{};
\node (e5) [below of=a32]{};
\node (d4) [right of=e4] {$G^{\vec n}$};
\node (d5) [right of=e5] {$G^{\vec n}$};
\draw[->](a31) -- node[above]{$W^\pm$} (d4);
\draw[->](a32) -- node[above]{$\gamma$}node[below]{$Z$} (d5);
\node (e6) [below of=a33]{};
\node (e7) [below of=a34]{};
\node (d6) [right of=e6] {$G^{\vec n}$};
\node (d7) [right of=e7] {$G^{\vec n}$};
\draw[->](a33) -- node[above]{$e^\pm$}node[below]{$\nu$} (d6);
\draw[->](a34) -- node[above]{$e^\pm$}node[below]{$\nu$} (d7);

\draw[->](a1) -- node[above]{$q$} (a3);
\draw[->](a1) -- node[below]{$q$}(a4);
\draw[->](a3) -- node[above]{$q^\prime$} (a31);
\draw[->](a3) -- node[below]{$q$} (a32);
\draw[->](a31) -- node[above]{$(\nu,~e^\pm)$} (a33);
\draw[->](a32) -- node[above]{$(e^\mp,~\nu)$} (a34);
\draw[->](a33) -- node[above]{$(e^\pm,~\nu)$} (a35);
\draw[->](a34) -- node[above]{$(e^\pm,~\nu)$} (a36);
\draw[->](a4) -- node[below]{$q$} (a37);
\draw[->](a35) -- node[above]{$\gamma~{\rm or}~Z$} (a38);
\draw[->](a36) -- node[above]{$\gamma~{\rm or}~Z$} (a39);
\draw[->](a37) -- node[above]{$\gamma~{\rm or}~Z$} (a40);

 \end{tikzpicture}
  \caption{Decay cascade of level-1 gluon ($g_1$) for $N=2~{\rm and}~4$, where $G^{\vec n}\ni h^{\vec n},~A^{\vec n}~{\rm or}~\phi^{\vec n}$.}
  \label{cascade_KKCD}
 \end{center}
\end{figure}
\end{itemize}   

\subsection{Search for $\gamma\gamma+p_T\!\!\!\!\!\!/~~$ by the {\em ATLAS} collaboration}
After discussing different signal topologies, we can now move on to the collider searches for this scenario. In this article, we have only concentrated on the $\gamma\gamma+p_T\!\!\!\!\!\!/~~$ signature. Diphoton events with large missing transverse energy were recently analyzed by the {\em ATLAS} collaboration in Ref.~\cite{ATLAS}. The analysis of the {\em ATLAS} collaboration is based on the data collected by the {\em ATLAS} detector in proton-proton collision at $\sqrt s=7$ TeV with an integrated luminosity of 3.1 fb$^{-1}$. The observed $\gamma\gamma+p_T\!\!\!\!\!\!/~~$ data  is consistent with the SM background prediction. The absence of any excess of such events was then translated to an upper bound on the radius of compactification $R$ of { mUED} model with gravity mediated decays. The {\em ATLAS} bound on $R$ is based on the following assumptions.
\begin{itemize}
\item For calculating the signal events, {\em ATLAS} group considers the pair production of level-1 KK-gluon and quarks. Subsequently, the KK-gluon and quarks are allowed to decay via cascades involving other KK-matter particles until reaching the lightest KK-particle (namely, the $\gamma_1$) at the end of the decay chain. Finally, $\gamma_1$ decays into gravity excitations. Therefore, the analysis of the {\em ATLAS} group is based on the assumption that pair production of KK-quarks and gluon results into $\gamma\gamma+p_T\!\!\!\!\!\!/~~$ signature with 100\% effective branching fraction. From the discussion of the previous section, it is obvious that this assumption is justified for $N=6$ because, in this case, the { GMD} widths are several orders of magnitude smaller than the { KKCD} widths. However, for $N=2~{\rm and}~4$, { GMD} widths become comparable with the { KKCD} widths. As a result, for $N=2~{\rm and}~4$, pair production of KK-quarks and gluon does not give rise to $\gamma\gamma+p_T\!\!\!\!\!\!/~~$ signature with 100\% effective branching fraction. We have reanalyzed the {\em ATLAS} results for $N=2~{\rm and}~4$.
\item The {\em ATLAS} collaboration also assumes that $\gamma_1$ decays into gravity excitations in association with a photon with a 100\% branching fraction. However, it is important to note that in the framework of { mUED} model, $\gamma_1$ is not the level-1 excitation of the SM photon. As a result of radiative correction, the mixing between $W^{3}_\mu$ and $B_\mu$ is highly suppressed for non-zero KK-modes and thus, $\gamma_1$ is almost purely the level-1 KK-excitation of $B_\mu$. $\gamma_1$ has gravity induced coupling with both photon and SM $Z$-boson. Therefore, $\gamma_1$ can decay into a photon or $Z$-boson in association with a gravity excitation. We found that for $R^{-1}\sim 500$ GeV, the decay branching fraction of $\gamma_1$ in to a $Z$-boson is $\sim$ 22\% which is surely not a negligible number. In our analysis, we have included both the decay modes of $\gamma_1$.  
\end{itemize}  
\begin{figure}[t]
\begin{center}
\epsfig{file=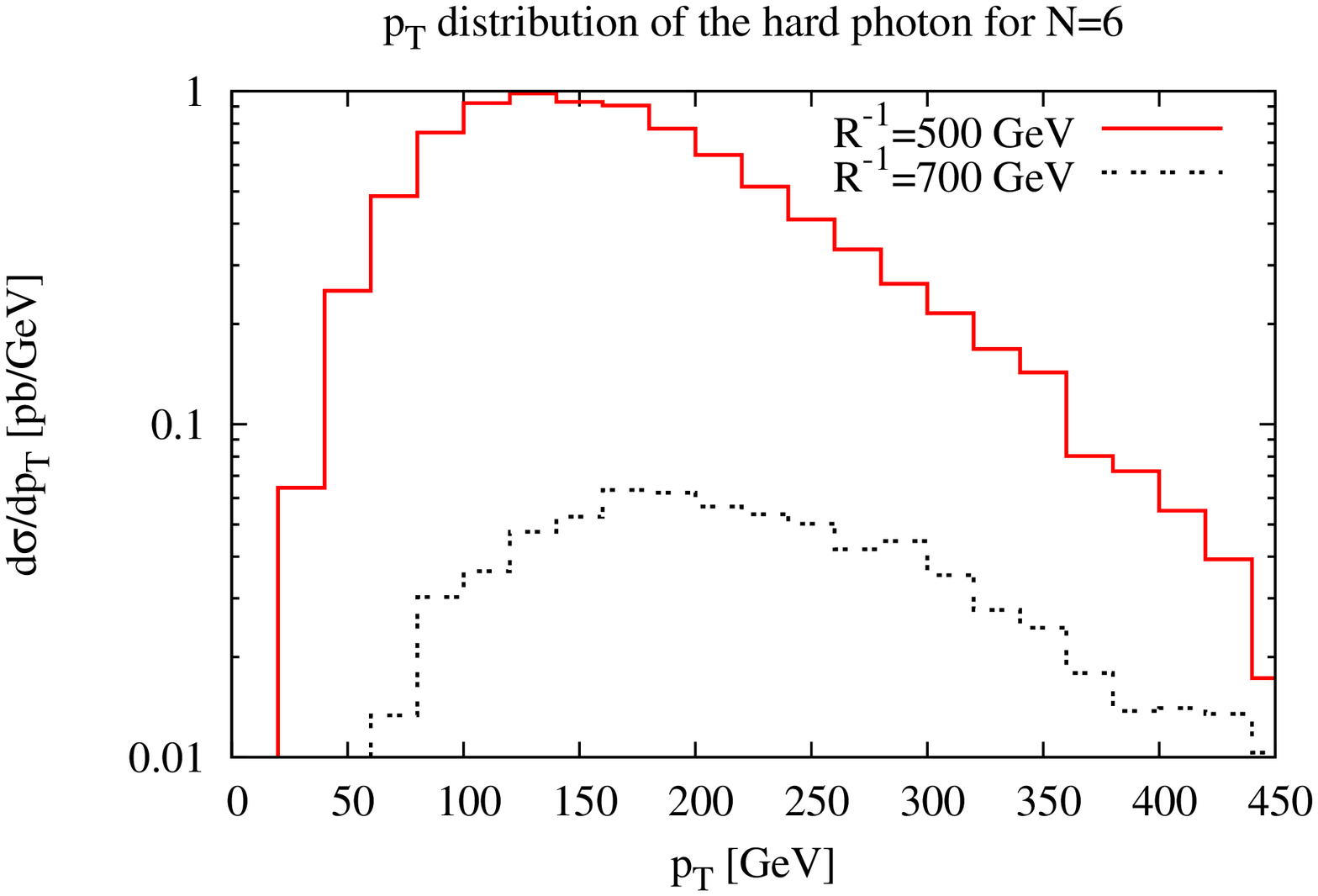,width=7cm,height=7cm,angle=270}
\epsfig{file=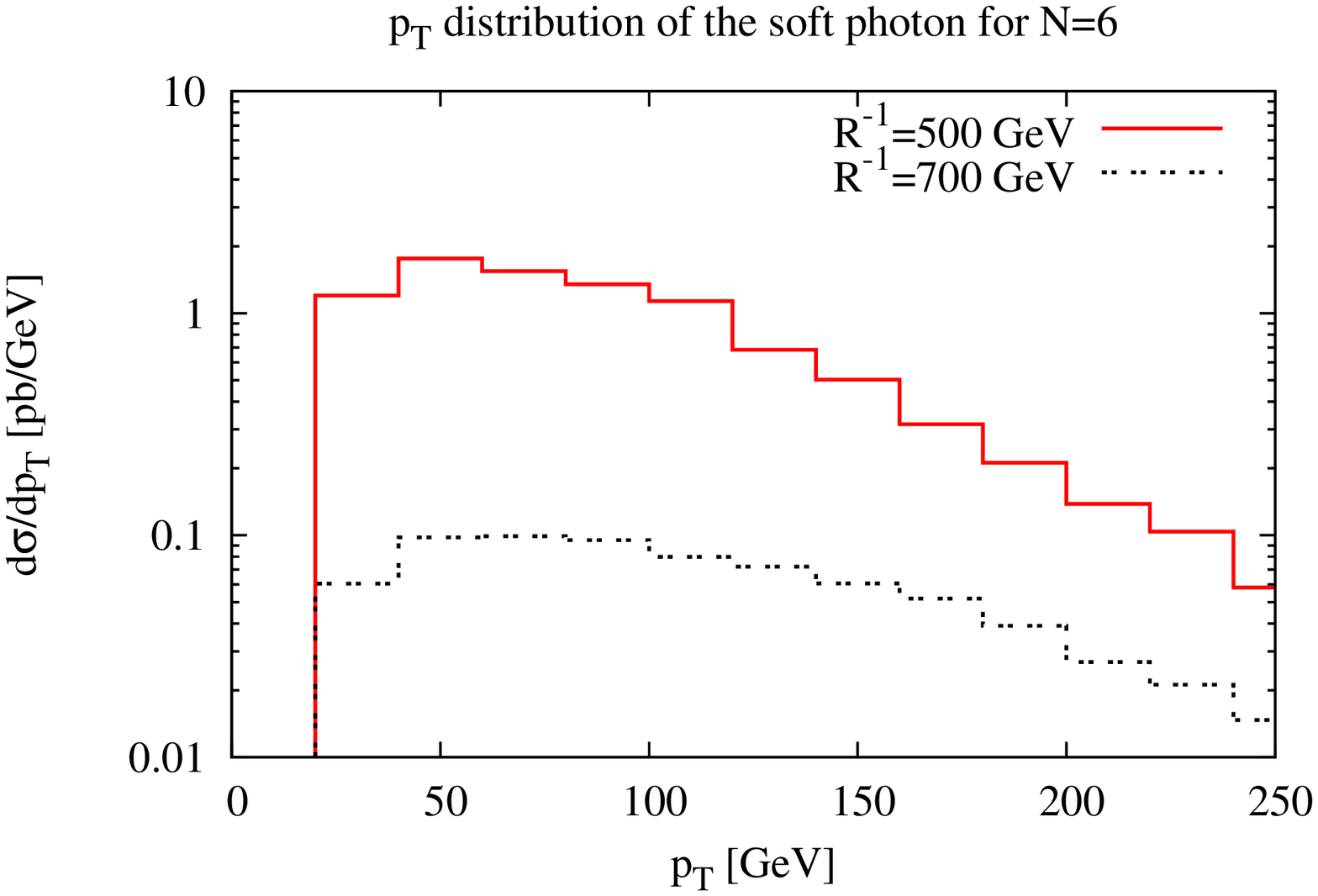,width=7cm,height=7cm,angle=270}
\end{center}
\caption{Transverse momentum distribution for the harder (left panel) and softer (right panel) photon (after ordering the photons according to their $p_T$ hardness) for $N=6$. In this plot, we have assumed the fundamental ($4+N$)D Planck mass $M_D=5$ TeV.}
\label{fig:gammapt}
\end{figure} 

Before going into the details of our analysis, we will briefly discuss about the analysis performed by the {\em ATLAS} collaboration. Events with at least two photons with $p_T>25 ~{\rm GeV}$, rapidity $|\eta| < 1.81$ and which are outside the transition region $1.37 < |\eta| < 1.52$ are analyzed by the {\em ATLAS} group. In addition, a photon isolation cut was applied, wherein the total energy deposit from all hadronic activity in a radius of $0.2$ in the $\eta-\phi$ space around the center of the photon had to be less than $35$ GeV. The reconstruction of $p_T\!\!\!\!\!\!/~~$ is based on the prescription in Ref.~\cite{ATLASmpt} with all topological calorimeter clusters within the rapidity coverage $|\eta|<4.5$. 

There are several sources of the SM background for the $\gamma \gamma + p_T\!\!\!\!\!\!/~~$ signal. The main source of the SM background is $\gamma \gamma$ production. Multi-jet and $\gamma+{\rm jets}$ events also contribute to the background if at least one jet is misidentified as a photon. Since QCD multi-jet production cross-section is huge, dominant contribution to the SM background results from multi-jet and $\gamma+{\rm jets}$ production. Production of the $W$-boson can also contribute to the background if $W$ decays leptonically ($W\to e\nu$) and the electron is misidentified as a photon. The second photon is either a real photon in $W \gamma$ events or a jet faking a photon in $W + {\rm jets}$ events. However, this contribution is very small compared to the QCD backgrounds.
 For the {\em ATLAS} analysis, the backgrounds were evaluated entirely using data. An independent { "misidentified jet"} control
sample, enriched in events with jets misidentified as photons, is used to model the $p_T\!\!\!\!\!\!/~~$ response for events with jets faking photons. The $p_T\!\!\!\!\!\!/~~$ response for the $\gamma \gamma$ events was modeled using the $p_T\!\!\!\!\!\!/~~$ spectrum measured in a high purity sample of $Z \to ee$ events. The $p_T\!\!\!\!\!\!/~~$ spectrum of the total background ($\gamma \gamma$, $\gamma+{\rm jets}$ and multi-jet) was modeled by a weighted sum of the $p_T\!\!\!\!\!\!/~~$ spectra of the $Z \to ee$ and { "misidentified jet"} samples.

For generating { BSM} ({mUED} with gravity mediated decay) signal events, {\em ATLAS} group used PYTHIA 6.421 \cite{PYTHIA} with the MC09 \cite{MC09} parameter tune. The radius of compactification, $R$,  was treated as the only free parameter for the { BSM} scenario. The number of { "large"} extra dimension ($N$) and the value of the ($4+N$)D Planck mass ($M_D$) were kept fixed at $N=6$ and $M_D=5$ TeV, respectively. The $p_T\!\!\!\!\!\!/~~$ spectrum of the observed $\gamma\gamma$ events is then compared with the total SM background prediction and { BSM} signal for different values of $R^{-1}$. The {\em ATLAS} collaboration found that the QCD background dominates in the low $p_T\!\!\!\!\!\!/~~$ region and falls sharply with increasing $p_T\!\!\!\!\!\!/~~$, whereas, the {mUED} signal is prominent at higher $p_T\!\!\!\!\!\!/~~$. The observed $p_T\!\!\!\!\!\!/~~$ spectrum of diphoton events is consistent with the predicted background over the entire $p_T\!\!\!\!\!\!/~~$ range. Since the { BSM} signal is expected to peak at higher $p_T\!\!\!\!\!\!/~~$, the signal search region was chosen to be $p_T\!\!\!\!\!\!/~~>75$ GeV. In the signal region, zero signal events were observed which is in good agreement with the SM background prediction. 
\begin{figure}[t]
\begin{center}
\epsfig{file=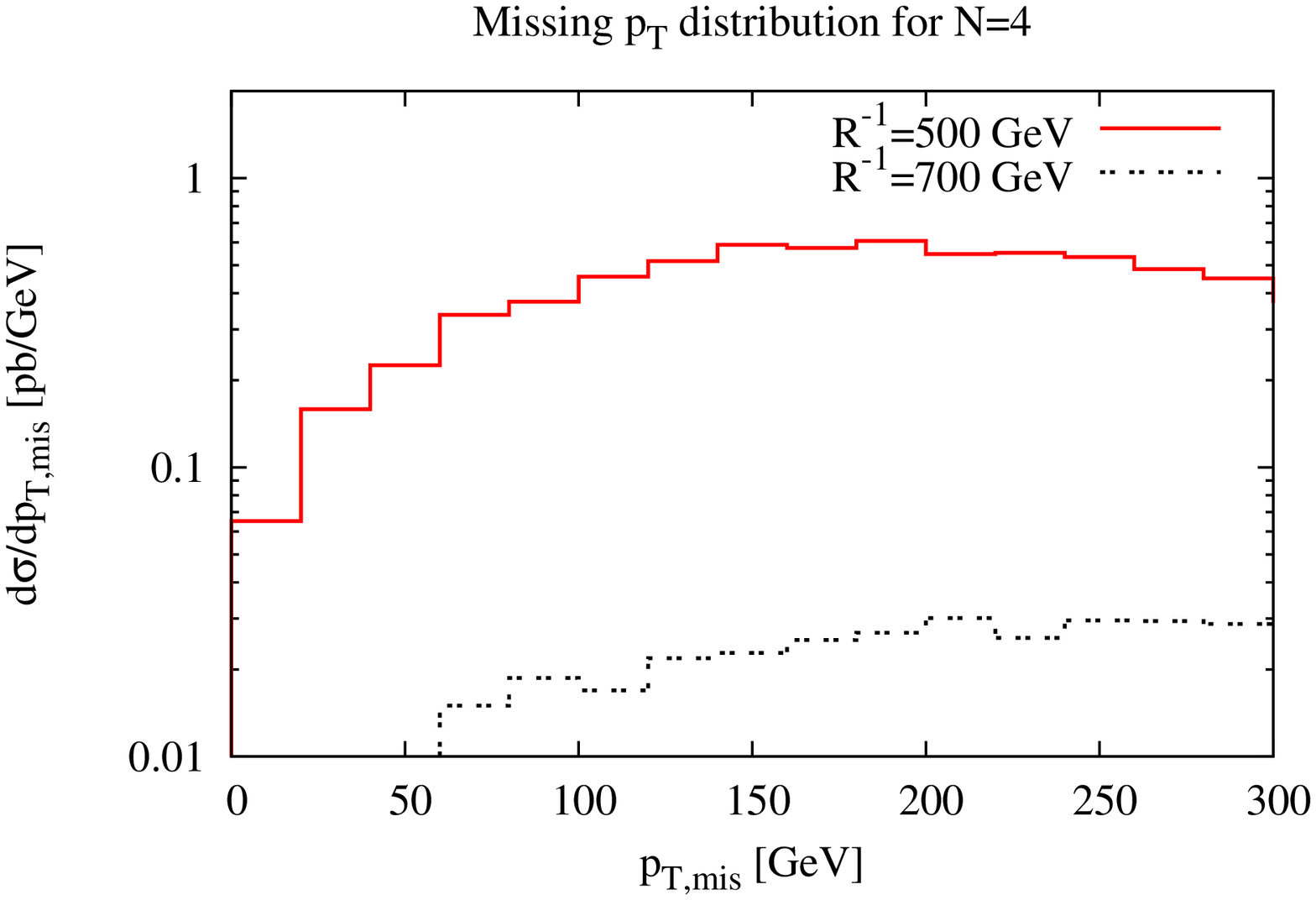,width=7cm,height=7cm,angle=270}
\epsfig{file=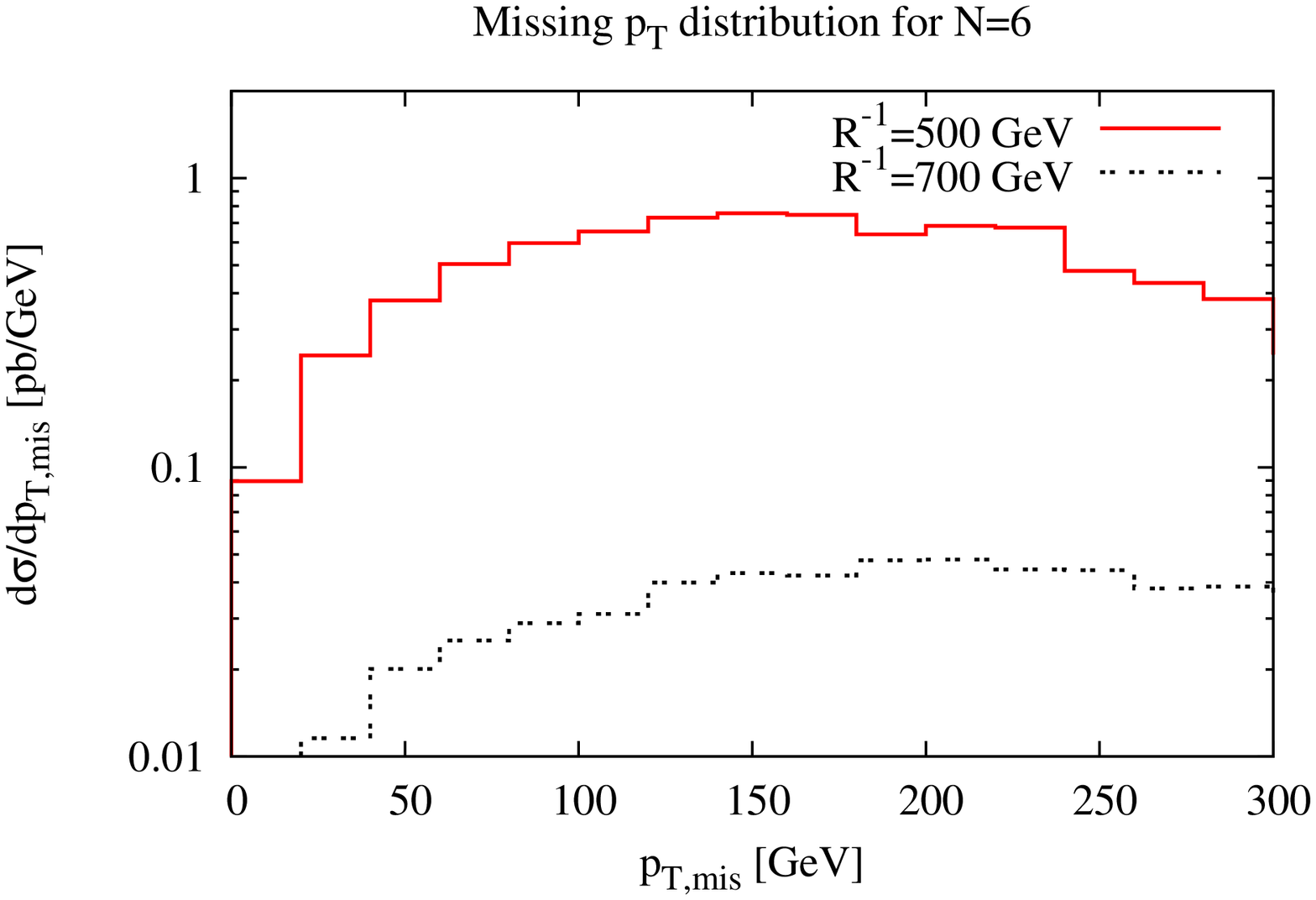,width=7cm,height=7cm,angle=270}
\end{center}
\caption{Missing transverse momentum distribution for $N=4$ (left panel) and $N=6$ (right panel). In this plot, we have assumed the fundamental ($4+N$)D Planck mass $M_D=5$ TeV.}
\label{fig:missingpt}
\end{figure} 

In our analysis, we have only estimated the signal diphoton cross-sections. Since the {\em ATLAS} group computed the background entirely from the data, the background computation is beyond the scope of this article. Nevertheless, to show the consistency of our calculation, we have computed few signal distributions and signal cross-sections for different $p_T\!\!\!\!\!\!/~~$ range. For the event generation and event selection, we have followed the prescription of {\em ATLAS} collaboration in Ref.~\cite{ATLAS}. The event generation technique and event selection criteria, used for our analysis, are summarized in the following:  
\begin{itemize}
\item {\bf Event Generation:} For generating the signal events, we have used PYTHIA 6.421 \cite{PYTHIA} with the implementation of the { mUED} model \cite{PYTHIAmUED}. However, the implementation of { mUED} in PYTHIA includes the gravity mediated decay for $\gamma_1$ only. We have modified the subroutine PYWIDT in PYTHIA to include the gravity mediated decays of the other level-1 KK-particles. For the tuning different PYTHIA parameters, we have used the MC09 parameter tune \cite{MC09}.
\item {\bf Event Selection:} We have selected events with at least two photons with transverse momenta, $p_T\ge 25$ GeV.  Both the photon candidates were required to have rapidity $|\eta| < 1.81$, and to be outside the transition region $1.37 < |\eta| < 1.52$ between the barrel and the end-cap calorimeters. Photons are required to be well isolated, {\em i.e.} the total energy deposit from all {\it hadronic activity} within a cone of radius $0.2$ around the lepton axis should be $\leq$ 35 GeV.  The missing transverse momentum ($p_T\!\!\!\!\!\!/~~$) in an event is computed by
\begin{equation}
 p_T\!\!\!\!\!\!/~~=\sqrt{\left(\sum p_x \right)^2+\left(\sum p_y\right)^2}. 
\end{equation}
Here the sum goes over all the isolated leptons, the jets, as well as
the "unclustered" energy deposits. 
\end{itemize}

In Fig.~\ref{fig:gammapt}, we have presented the photon transverse momentum distributions (after ordering the photons according to their $p_T$ hardness) for $N=6$ and $M_D=5$ TeV. We have assumed two different values of $R^{-1}=500~{\rm and}~700$ GeV. In Fig.~\ref{fig:missingpt}, we have presented the $p_T\!\!\!\!\!\!/~~$ distributions for the { mUED} signal events for $N=4$ (left panel) and $6$ (right panel) with $M_D=5$ TeV. We have assumed two different values of $R^{-1}=500~{\rm and}~700$ GeV. As discussed in the previous paragraph, Fig.~\ref{fig:missingpt} shows that $p_T\!\!\!\!\!\!/~~$ distributions for the signal events peak at higher values. Moreover, Fig.~\ref{fig:missingpt} also shows that the shape of the $p_T\!\!\!\!\!\!/~~$ distributions are similar for $N=4$ (left panel) and $6$ (right panel). In Table~\ref{tab:signal}, we have presented the signal $\gamma\gamma+p_T\!\!\!\!\!\!/~~$ cross-sections after the event selection cuts in several $p_T\!\!\!\!\!\!/~~$ range. The cross-sections are presented for two different values of $R^{-1}$ and $N=6,~4~{\rm and} ~2$. To verify the consistency of our computation, we have also presented the cross-sections assuming $N=6$ and $\gamma_1$ decays into $\gamma G^{\vec n}$ with 100\% branching fraction (see Table~\ref{tab:signal}, 2nd and 3rd column). With this assumption, we find that our signal cross-sections are consistent with the number of signal events obtained by the {\em ATLAS} collaboration in Ref.~\cite{ATLAS}. Table~\ref{tab:signal} shows that for $N=2$, signal cross-sections are significantly smaller compared to $N=4~{\rm and}~6$. This can be attributed to the fact that for $N=2$, { gravity mediated decays} of KK-matter fields become dominant over the { KK-number conserving decays} and thus, the effective branching fractions of the level-1 KK-particles into $\gamma G^{\vec n}$ pairs gets suppressed. 

\begin{table}[h]
\begin{center}
\begin{tabular}{||c||c|c||c|c||c|c||c|c||}
\hline \hline
 & \multicolumn{8}{|c||}{Signal Cross-Section in pb}\\\cline{2-9}
Missing& \multicolumn{2}{|c||}{When $\gamma_1\to \gamma G^{\vec n}$} & \multicolumn{6}{|c||}{When $\gamma_1\to \gamma G^{\vec n}$}\\
 $p_T$  & \multicolumn{2}{|c||}{with 100\% BF}  & \multicolumn{6}{|c||}{and $\gamma_1\to Z G^{\vec n}$}\\\cline{2-9}
range   & \multicolumn{2}{|c||}{$N=6$} & \multicolumn{2}{|c||}{$N=6$}  &  \multicolumn{2}{|c||}{$N=4$} &  \multicolumn{2}{|c||}{$N=2$} \\\cline{2-9}
 in GeV & \multicolumn{2}{|c||}{$R^{-1}$ in GeV} & \multicolumn{2}{|c||}{$R^{-1}$ in GeV} &\multicolumn{2}{|c||}{$R^{-1}$ in GeV} & \multicolumn{2}{|c||}{$R^{-1}$ in GeV} \\\cline{2-9}
 & 500 & 700 & 500 & 700 & 500 & 700 & 300 & 500 \\\hline\hline
$20 - 30$ & 0.16 & $7.2\times 10^{-3}$ & $0.09$ & $3.8\times 10^{-3}$ &$0.05$ & $2.5\times 10^{-3}$ &0.58 & $1.2\times 10^{-3}$\\
$30 - 50$ &  0.48 & $2.2\times 10^{-2}$ & 0.28 & $1.7\times 10^{-2}$ & 0.22 &$8.3\times 10^{-3}$ & 1.75 &$1.9\times 10^{-3}$\\
$50 - 75$ &  0.88 &$4.2\times 10^{-2}$ & 0.57 & $3.0\times 10^{-2}$ & 0.38 &$1.6\times 10^{-2}$ & 2.08 & $8.9\times 10^{-3}$\\
$\ge 75$ & 12.6 & 1.35 & 7.98 & 0.75 & 7.87 & 0.6 & 29.5 & 0.21\\\hline\hline

\end{tabular}

\end{center}
\caption{{\rm mUED} with gravity mediated decay scenario contribution $\gamma\gamma+p_T\!\!\!\!\!\!/~~$ cross-sections (in pb) for different $p_T\!\!\!\!\!\!/~~$ range after the event selection cuts. The cross-sections are presented for $M_D=5$ TeV and three different values of $N=6,~4~{\rm and}~2$. To check the consistency of our calculation, in 2nd and 3rd column, we present the signal cross-sections for $N=6$ with $\gamma_1$ decaying only into $\gamma G^{\vec n}$.}  
\label{tab:signal}
\end{table}
\begin{figure}[t]
\begin{center}
\epsfig{file=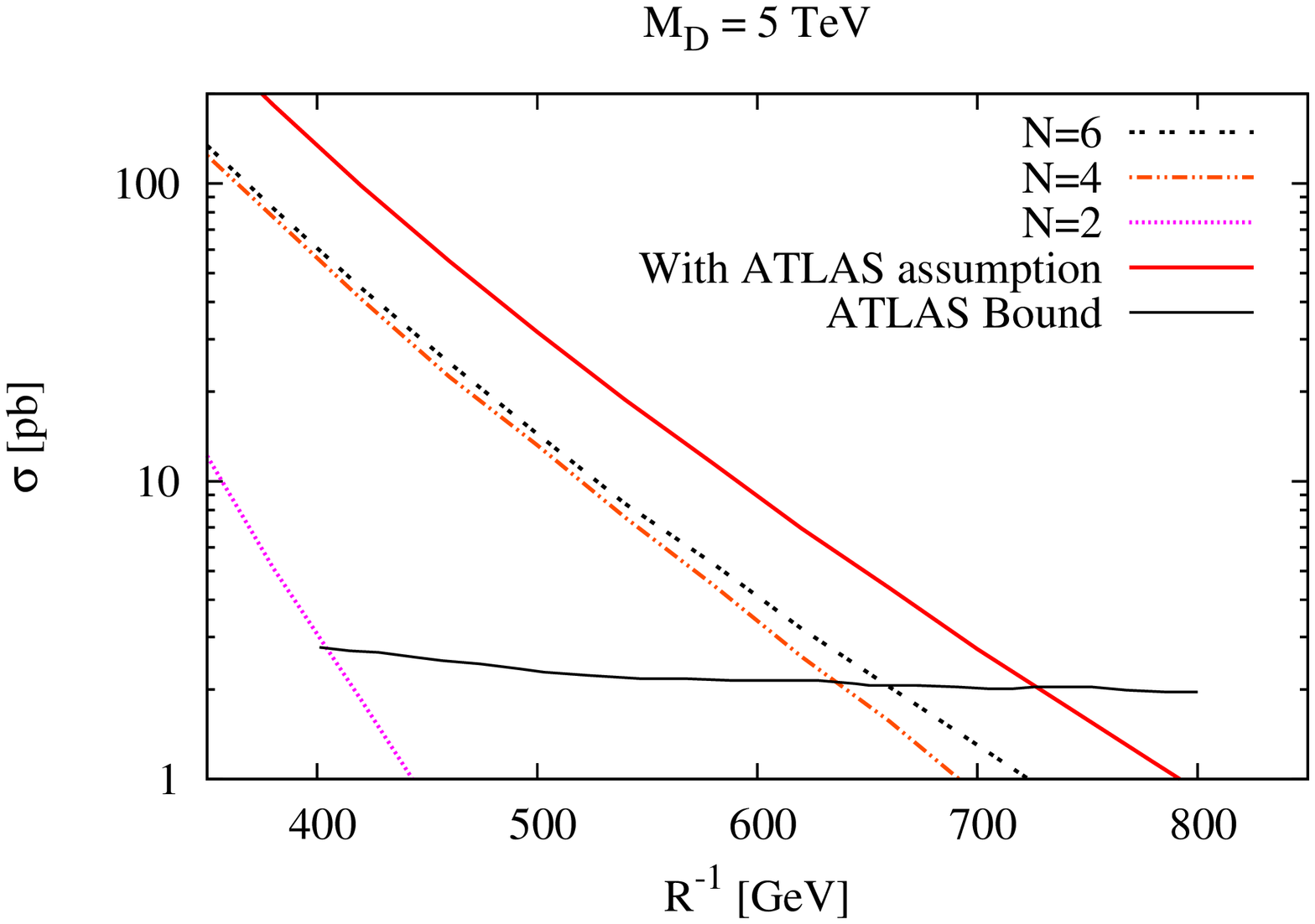,width=9cm,height=9cm,angle=270}
\end{center}
\caption{Black line corresponds to 95\% CL upper limits on the theory $\gamma\gamma+p_T\!\!\!\!\!\!/~~$ production cross-section obtained by the {\em ATLAS} group \cite{ATLAS}. Other lines correspond to the LO theory prediction for $\gamma\gamma+p_T\!\!\!\!\!\!/~~$ cross-section for $N=2,~4~{\rm and}~6$. To check the consistency of our computation, we also present the theory cross-section computed with the {\em ATLAS} assumption i.e., $N=6$ and  BF$(\gamma_1\to \gamma G^{\vec n})=100\%$.
}
\label{fig:reach}
\end{figure} 

\subsection{Exclusion limits on the model parameters}
\begin{figure}[t]
\begin{center}
\epsfig{file=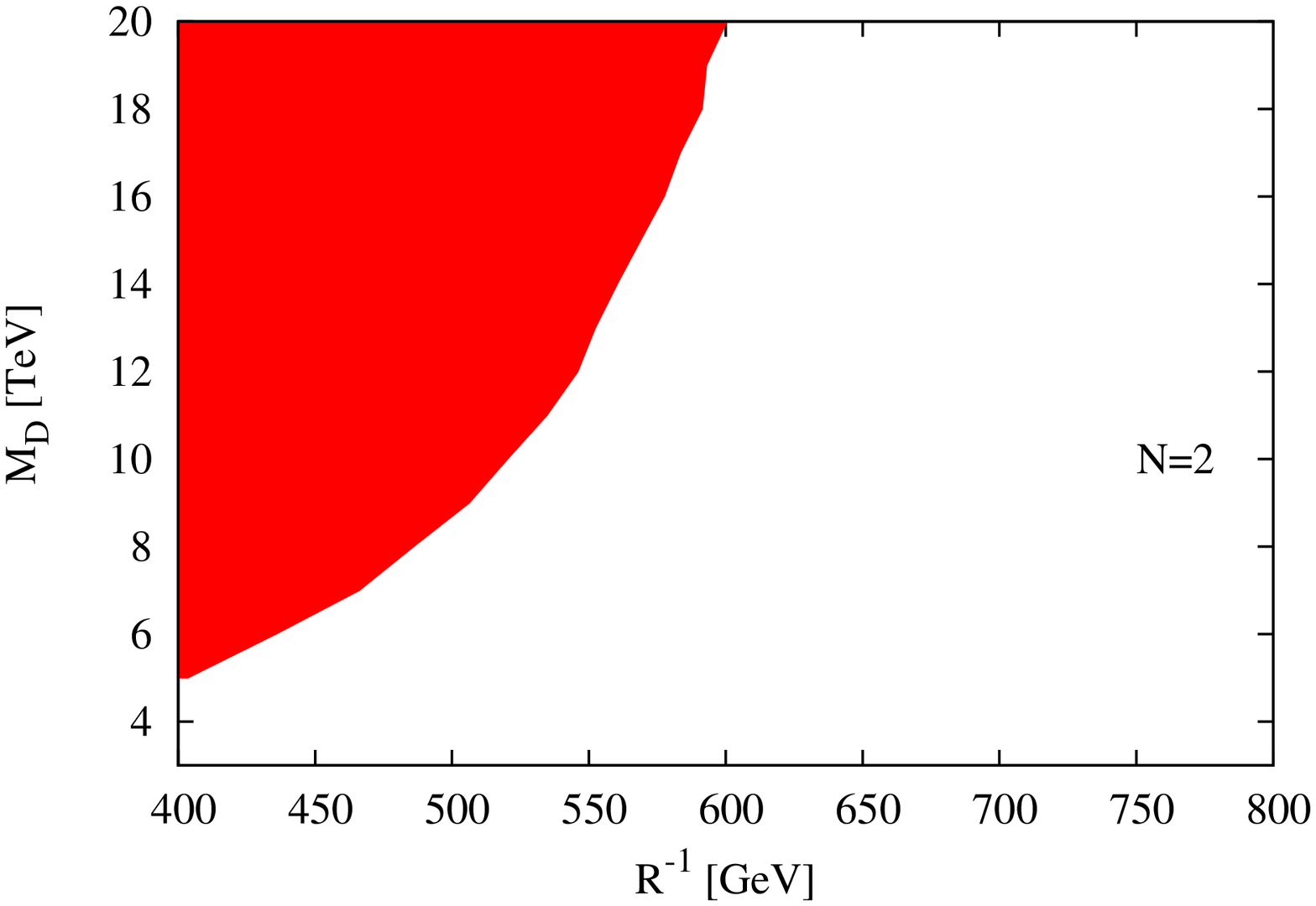,width=7cm,height=7cm,angle=270}
\epsfig{file=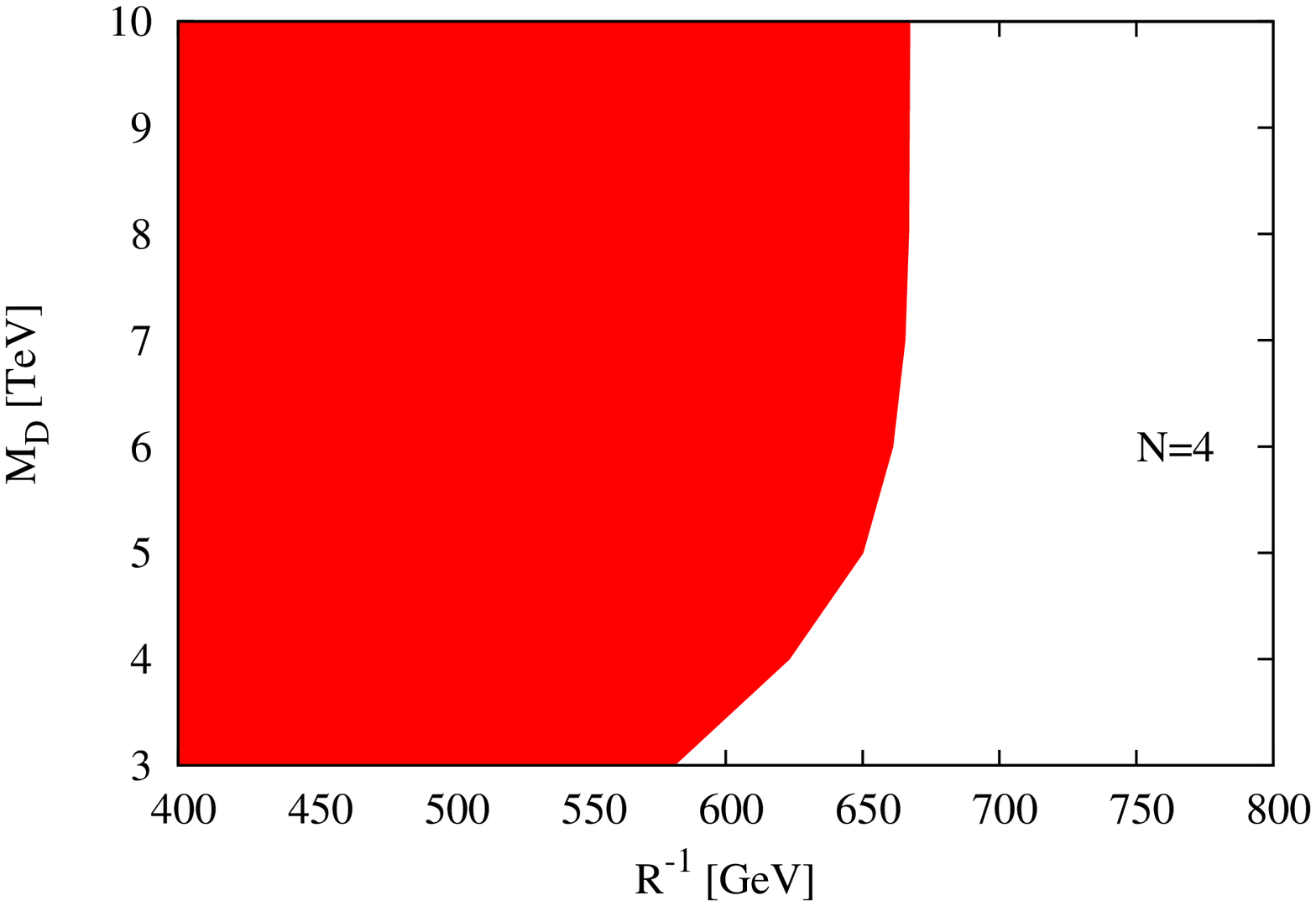,width=7cm,height=7cm,angle=270}\\
\epsfig{file=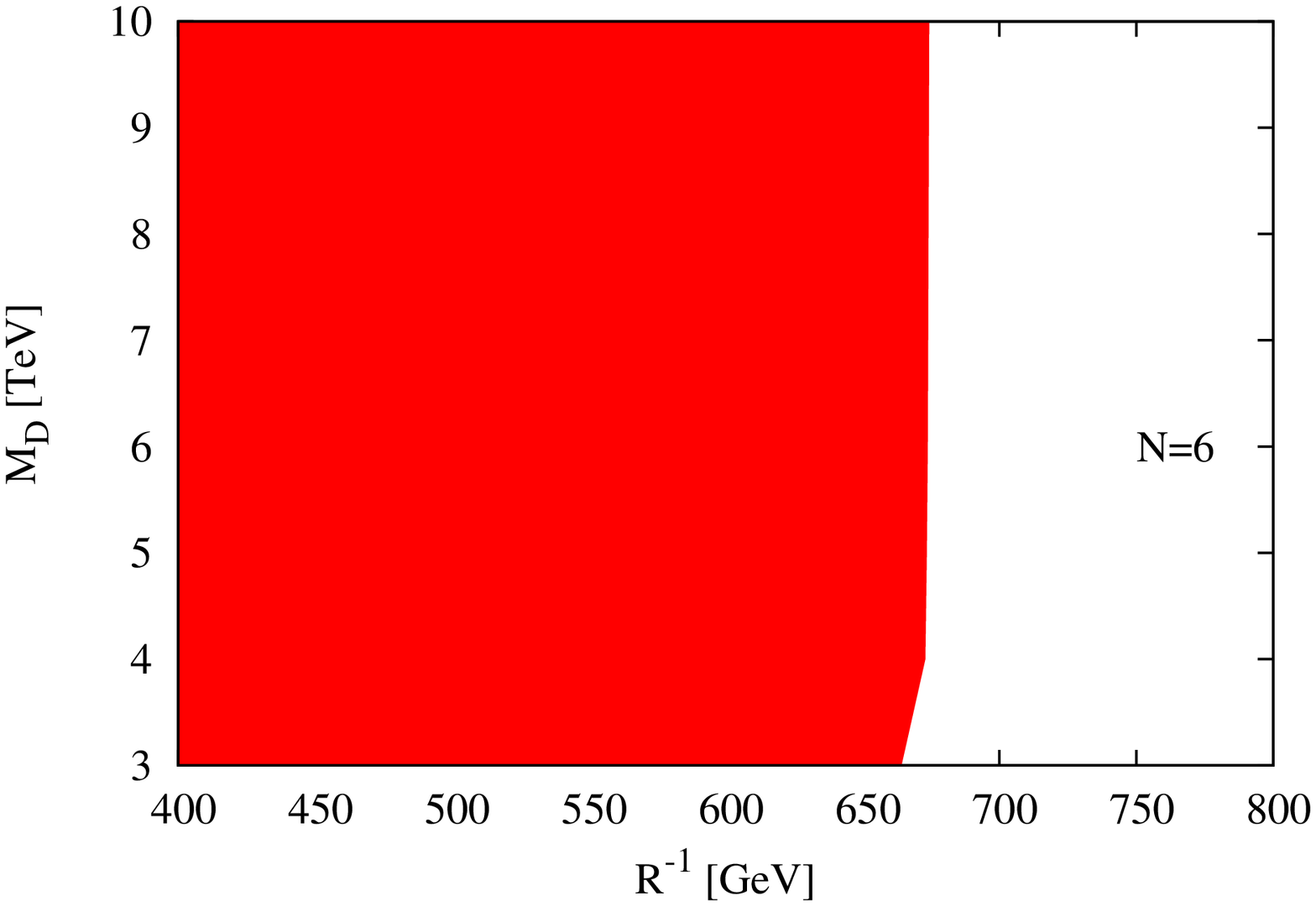,width=7cm,height=7cm,angle=270}
\end{center}
\caption{95\% CL excluded parameter space of { mUED} model with gravity mediated decay for $N=2$ (top left panel), $4$ (top right panel) and $6$ (bottom panel). The dark region in the $R^{-1}-M_D$ plane is excluded at 95\% CL from {\em ATLAS} diphoton search.}
\label{fig:scan}
\end{figure} 

The diphoton data observed by the {\em ATLAS} detector is in good agreement with the SM background prediction over the entire $p_T\!\!\!\!\!\!/~~$ range. { mUED} model with { gravity induced decays} of the KK-particles gives rise to the diphoton signal with large $p_T\!\!\!\!\!\!/~~$. However, there is no indication of an excess of diphoton events in the high $p_T\!\!\!\!\!\!/~~$ region. As for example, in the { signal search region} (defined by $p_T\!\!\!\!\!\!/~~>75$ GeV), there are zero observed events which is consistent with the predicted SM background. Whereas, Table~\ref{tab:signal} shows that if { mUED} with gravity mediated decay scenario with $R^{-1}\sim 500$ GeV and $N=4~(6)$ exists in nature then the {\em ATLAS} detector should have observed $\sim$ 24 (25) diphoton events with $p_T\!\!\!\!\!\!/~~>75$ GeV at an integrated luminosity $3.1~{\rm pb}^{-1}$. As a result, $R^{-1}=500$ GeV for $N=4~(6)$ is excluded. 

In view of the good agreement between observed $p_T\!\!\!\!\!\!/~~$ distribution and expected SM background $p_T\!\!\!\!\!\!/~~$ distribution, an 95\% CL upper limit on the total { mUED} production cross-section multiplied by the diphotonic effective { Branching Fraction (BF)}\footnote{In Ref.~\cite{ATLAS}, the 95\% CL upper limit was set on the total { mUED} production cross-section as a function of $R^{-1}$ assuming that the pair production of KK-particles gives rise to diphoton signal with 100\% { BF} fraction. This assumption is true for $N=6$ and BF($\gamma_1\to \gamma G^{\vec n}$)=100\%. In general, the { mUED} production cross-section which contributes to the diphoton signal, is given by the total { mUED} production cross-section multiplied by the effective diphotonic BF. Therefore, we can generalize the ATLAS 95\% CL limit as the upper limit on the total { mUED} production cross-section multiplied by the effective diphotonic BF.} was set by ATLAS group. The upper limit is based on the number of observed and expected events with $p_T\!\!\!\!\!\!/~~>75$ GeV. In Fig.~\ref{fig:missingpt}, we have shown that the shape of the $p_T\!\!\!\!\!\!/~~$ distribution does not significantly depend on the number of { "large"} extra dimensions.  Therefore, the cross-section upper limit which was obtained by the ATLAS group for $N=6$, is also applicable for $N=2~{\rm and}~4$. In Fig.~\ref{fig:reach}, we have presented the 95\% CL upper limit obtained by the ATLAS group together with the theory contribution to the diphoton cross-section as a function of $R^{-1}$ for $M_D=5$ TeV. The theory cross-sections are presented for $N=2,~4~{\rm and}~6$. To show the consistency of our computation, in Fig.~\ref{fig:reach}, we also present theory cross-section, computed with the ATLAS assumptions, {\em i.e.} $N=6$ and BF($\gamma_1\to \gamma G^{\vec n}$)=100\%. Fig.~\ref{fig:reach} shows that the lower bounds on $R^{-1}$ depend on the number of { "large"} extra dimension. As for example, for $M_D=5$ TeV, $R^{-1}$ is ruled out upto 403, 650 and 673 GeV for $N=2,~4~{\rm and}~6$ respectively. Here, it is important to mention that the { gravity mediated decay} widths and hence, the { mUED} diphoton cross-section, depend on both $R^{-1}$ and $M_D$. Therefore, the lower bounds on the $R^{-1}$ should also depend on $M_D$. To constrain the parameters $R^{-1}$ and $M_D$, we have scanned over the $R^{-1}-M_D$ plane. In Fig.~\ref{fig:scan}, we have shown the region in $R^{-1}-M_D$ plane which is excluded by the ATLAS diphoton search with 95\% CL. Fig.~\ref{fig:scan} shows that for $N=2$, $R^{-1}\le 400~(600)$ GeV region is ruled out for $M_D=5~(20)$ TeV. Whereas, for $N=4$, $R^{-1}$ is excluded upto 575 (670) GeV for $M_D=3~(10)$ TeV.

\section{Conclusion}

To summarize, we have discussed the signature of { fat brane} scenario in the context of the LHC experiment with $\sqrt s=7$ TeV and integrated luminosity 3.1 pb$^{-1}$. In particular, we have concentrated on the { diphoton}$+p_T\!\!\!\!\!\!/~~$ signature. In the framework of { fat brane} scenario, apart from the { KK-number conserving } decays into lighter KK-particles, level-1 KK-particles also decay into corresponding SM particle by radiating gravity excitations. The pair production of level-1 particles at the LHC gives rise to different final state signal topologies. In this article, we have concentrated on the { diphoton}$+p_T\!\!\!\!\!\!/~~$ signature. Conservation of { KK-parity} allows the pair production of level-1 KK-particles only. In the framework of this model, { diphoton}$+p_T\!\!\!\!\!\!/~~$ signature results when produced the pair of level-1 KK-particles decay via cascades involving other KK-particles until reaching the LKP and the LKP further decays into a photon and gravity excitation.

Diphoton events with large missing transverse energy were studied by the {\em ATLAS} collaboration for proton-proton collisions at $\sqrt s=7$ TeV and integrated luminosity $3.1$pb$^{-1}$. The observed diphoton data was consistent with the SM background prediction. In this work, we have translated this {\em ATLAS} result to constrain the parameter space of this model. We find that the inverse radius of compactification ($R^{-1}$) upto 650 (673) GeV can be ruled out for $M_D=5$ TeV in the case of $N=4~(6)$, whereas, for $N=2$, {\em ATLAS} result implies that $R^{-1}$ is excluded only upto 403 GeV.

\newpage
\appendix
 \renewcommand{\theequation}{A-\arabic{equation}}
  \setcounter{equation}{0}  

\section*{Appendix}
\section{{\em Kaluza-Klein} expansion of the fields}
\label{KKEXP}
 Defining
\beq
{\cal C}_n \equiv \sqrt{\frac{2}{\pi \, R}} \, \cos \, \frac{n \, y}{R} 
\qquad 
{\rm and}
\qquad
{\cal S}_n \equiv \sqrt{\frac{2}{\pi \, R}} \, \sin \, \frac{n \, y}{R} \ ,
\eeq
the KK-expansions are given by
\beq
\label{fourier}
\barr{rcl}
A_{\mu}(x,y)&=&\dis \frac{1}{\sqrt{2}} \, A_{\mu}^{(0)}(x) \, {\cal C}_0 
               + \sum^{\infty}_{n=1} A_{\mu}^{(n)}(x) \, {\cal C}_n \ ,
\\[1ex]
A_5(x,y) & = & \dis \sum^{\infty}_{n=1}A_5^{(n)}(x)\, {\cal S}_n  \ ,
\\[1ex]
\phi(x,y)&=& \dis \frac{1}{\sqrt{2}} \, \phi^{(0)}(x) {\cal C}_0 
    + \sum^{\infty}_{n=1}\phi^{(n)}(x)  \, {\cal C}_n \ ,
\\[2.5ex]
Q_{i}(x,y)&=&\dis \frac{1}{\sqrt{2}} \, Q_{iL}^{(0)}(x) \,
                             {\cal C}_0  
    + \sum^{\infty}_{n=1}\left[ Q^{(n)}_{iL}(x) \, {\cal C}_n +
                        Q^{(n)}_{iR}(x) \, {\cal S}_n \right] \ ,
 \\[2.5ex]
u_{i}(x,y)&=&\dis \frac{1}{\sqrt{2}} \, u_{iR}^{(0)}(x) \,
                             {\cal C}_0  
    + \sum^{\infty}_{n=1}\left[ u^{(n)}_{iR}(x)\, {\cal C}_n
                       + u^{(n)}_{iL}(x)\, {\cal S}_n \right] \ ,
 \\[2.5ex]
d_{i}(x,y)&=&\dis \frac{1}{\sqrt{2}} \, d_{iR}^{(0)}(x) \,
                             {\cal C}_0  
    + \sum^{\infty}_{n=1}\left[ d^{(n)}_{iR}(x)\, {\cal C}_n
                        + d^{(n)}_{iL}(x)\, {\cal S}_n \right] \ ,
\earr
    \label{mode_expan}
\eeq
where $i=1\dots3$ denotes generations and the fields $Q_i$,
$u_i$, and $d_i$ describe the 5-dimensional quark
weak-doublet and singlet states respectively.  The zero modes thereof
are identified with the 4-dimensional chiral SM quark states.  The
complex scalar field $\phi (x,y)$ and the gauge boson $A_\mu(x,y)$ are
$Z_2$ even fields with their zero modes identified with the SM scalar
doublet and SM gauge bosons respectively. On the contrary, the field
$A_5(x,y)$, which is a real scalar transforming in the adjoint
representation of the gauge group, does not have any zero mode.  The
KK-expansions of the lepton fields are analogous to those for the quarks
and are not shown for the sake of brevity.

\section{Radiative corrections to the KK-masses}
\label{correction}

\begin{itemize}
\item Bulk corrections:\\
These arise
due to the winding of the internal loop (lines) around the compactified
direction\cite{Cheng:2002iz}, 
and are nonzero (and finite) only for the gauge boson KK-excitations.
For the first level KK-modes, the bulk corrections are given by  
\beq
\barr{rcl}
\delta\, (m_{B^n}^2) &=& \dis -\frac{39 \,\zeta(3) \, a_1}{2 \, \pi^2 \, R^2} \ ,
 \\[2ex]
\delta\, (m_{W^n}^2) &=&  \dis  -\frac{5\,\zeta(3) \, a_2}{2 \, \pi^2 \, R^2}\ ,
\\[2ex]
\delta\, (m_{g^n}^2) &=& \dis  -\frac{3\,\zeta(3) \, a_3}{2 \, \pi^2 \, R^2}\ ,
\earr
\label{delta0}
\eeq
where $a_i \equiv g_i^2 / 16 \, \pi^2 \ , i = 1\dots 3$ with $g_i$ 
denoting the respective gauge coupling constants. 
The vanishing of these corrections in the limit $R \rightarrow \infty$
reflects 
the removal of the compactness of the fifth direction and hence the 
restoration of full five-dimensional Lorentz invariance.

\item Orbifold corrections:\\ 
The very process of orbifolding introduces a set of special (fixed)
points in the fifth direction (two in the case of $S^1/Z_2$
compactification). This clearly violates the five-dimensional Lorentz
invariance of the tree level Lagrangian.  Unlike the bulk corrections,
the boundary corrections are not finite, but are logarithmically
divergent\cite{Cheng:2002iz}.  They are just the counterterms of the
total orbifold correction, with the finite parts being completely
unknown, dependent as they are on the details of the ultraviolet
completion.  Assuming that the boundary kinetic terms vanish at the cutoff
scale $\Lambda$ ($=20$ TeV here) the corrections from the boundary terms,
at a renormalization scale $\mu$ would obviously be proportional to 
$L_0 \equiv \ln (\Lambda^2 /\mu^2) $. Denoting $m_n({\cal A})$ to be the 
tree-level mass of the $n$-th KK-component of a SM field ${\cal A}$, we 
have\cite{Cheng:2002iz}

\beq
\barr{rcl}
\bar{\delta}\, m_{Q^n} &=& \dis 
m_n \,\left(3\, a_3
+ \frac{27}{16}\,a_2 + \frac{a_1}{16}
\right) \,L_0 \ ,
\\[2ex]
\bar{\delta}\, m_{u^n} &=& m_n \,\left(3\, a_3
  + a_1
\right) \, L_0 \ ,
\\[2ex]
\bar{\delta}\, m_{d^n} &=& \dis m_n \,\left(3\, a_3
  + \frac{a_1}{4}
\right) \, L_0 \ ,  \\[2ex]
\bar{\delta}\, m_{L^n} &=& \dis m_n \,\left(
 \frac{27}{16}\,a_2 + \frac{9}{16}\,a_1
\right) \, L_0 \ ,  \\[2ex]
\bar{\delta}\, m_{e^n} &=& \dis
 \frac{9\,a_1}{4}
\,  m_n \,L_0 \ , \\[2ex]
\bar{\delta}\, (m_{B^n}^2) &=& \dis \frac{-a_1}{6}\, m_n^2\,  L_0 
 \\[2ex]
\bar{\delta}\, (m_{W^n}^2) &=& \dis \frac{15\, a_2}{2}
\, m_n^2\, L_0 \ , \\[2ex]
\bar{\delta}\, (m_{g^n}^2) &=& \dis \frac{23\, a_3}{2} \,  m_n^2 \,
L_0 \ , \\[2ex]
\bar{\delta}\, (m_{H^n}^2) &=& \dis m_n^2 \, \left(\frac{3}{2}\, a_2   
+ \frac{3}{4}\, a_1 - \frac{\lambda_H}{16\pi^2} \right)
L_0 + \overline{m}_H^2\ , 
\earr
\label{delta1}
\eeq
where the boundary term for the Higgs scalar, namely $\overline{m}_H^2$,
 is taken to be vanishing. 
\end{itemize}

\end{document}